\documentclass[10pt,aps,prx,twocolumn,notitlepage,showpacs,superscriptaddress]{revtex4-2}

\usepackage{graphicx}
\usepackage{amssymb,amsmath,bm}
\usepackage{footnote}
\usepackage[colorlinks, linkcolor=blue,anchorcolor=blue,citecolor=blue,urlcolor=blue]{hyperref}
\newcommand{\be}{\begin{equation}}
\newcommand{\ee}{\end{equation}}

\renewcommand{\b}[1]{{\boldsymbol{#1}}}

\newcommand{\sgn}{\mathop{\mathrm{sgn}}}

\begin{document}

\title{
 Spin susceptibility in interacting two-dimensional semiconductors and bilayer systems at first order: Kohn anomalies and spin density wave ordering}

\author{Joel Hutchinson}
\affiliation{Department of Physics, University of Basel, Klingelbergstrasse 82, CH-4056 Basel, Switzerland} 

\author{Dmitry Miserev}
\affiliation{Department of Physics, University of Basel, Klingelbergstrasse 82, CH-4056 Basel, Switzerland} 

\author{Jelena Klinovaja }
\affiliation{Department of Physics, University of Basel, Klingelbergstrasse 82, CH-4056 Basel, Switzerland} 

\author{Daniel Loss}
\affiliation{Department of Physics, University of Basel, Klingelbergstrasse 82, CH-4056 Basel, Switzerland} 
\date{\today}

\begin{abstract}
This work is an analytic theoretical study of a 2D semiconductor with a Fermi surface that is split by the Zeeman coupling of electron spins to an external magnetic field in the presence of electron-electron interactions.
For the first time, we calculate the spin susceptibility for long-range and finite-range interactions diagrammatically, and find a resonant peak structure at the Kohn anomaly already in first-order perturbation theory. 
In contrast to the density-density correlator that is suppressed due to the large electrostatic energy required to stabilize charge density order, the spin susceptibility does not suffer from electrostatic screening effects, thus favouring spin-density-wave order in 2D semiconductors. 
Our results impose significant consequences for determining magnetic phases in 2D semiconductors.
For example, a strongly enhanced Kohn anomaly may result in helical ordering of magnetic impurities due to the RKKY interaction.
Furthermore, the spin degree of freedom can equally represent a layer pseudospin in the case of bilayer materials. 
In this case, the external ``magnetic field'' is a combination of layer bias and interlayer hopping. 
The sharp peak of the 2D static spin susceptibility may then be responsible for dipole-density-wave order in bilayer materials at large enough electron-phonon coupling.
\end{abstract}
\maketitle

\section{Introduction}

Two-dimensional (2D) semiconductors are increasingly popular materials in condensed matter physics due to various high-precision techniques, allowing physicists to control the chemical content, geometry, strain, disorder, electron density, and various spectral features of an experimental device.
The ability to tune the electron density in a 2D device by electrostatic gates enables  extensive experimental studies of a variety of phase transitions and emergent orders, from metal-to-insulator transitions and non-Fermi-liquid behavior \cite{mason,kravchenko,coleridge,simmons,papadakis,yoon,shashkin,keser,chowdhury,lyu,jaoui}, to spin and valley polarized ground states \cite{tongay,bonilla,roch1,roch2,hossain2020,ma}.

In this paper we study spin and charge density wave (DW) ordering in doped 2D semiconductors,
the phenomenon that has attracted much attention due to potential applications in electronics, photodetection, and information storage~\cite{hossain2017}. 
While in one-dimensional metals, DW order emerges due to long-lived collective spin and charge excitations~\cite{peierls1955,giamarchi,tsunetsugu,xavier,braunecker,japaridze,klinovajastano}, DW ordering in 2D metals is often associated with either hot-spot physics or perfectly nested Fermi surfaces (FS) \cite{abanov,hartnoll,allais,metzner}.
Instead, in this work we study the effect of the repulsive electron-electron interaction on spin and charge susceptibilities of an isotropic 2D electron gas (2DEG), which has neither hot spots nor perfect FS nesting.

Non-analyticities of a 2DEG have been mostly studied for a zero-range contact interaction \cite{lohneysen,vojta,belitzkirk1997,chubmaslov2003,chubglazman2005,chesizak,zak2012,maslov,zak2010,mis2019,mis2021,mis2022}, where the first non-analytic contributions emerge from second-order perturbation theory.
These non-analyticities do not provide resonant peak structures in static susceptibilities, yet they still could be responsible for first-order magnetic quantum phase transitions~\cite{roch2,lohneysen,vojta,belitzkirk1997,chubmaslov2003,chubglazman2005,chesizak,zak2012,maslov,zak2010,mis2019,mis2021,mis2022}.
Here, we show analytically for the first time that, in contrast to a zero-range interaction, any finite-range repulsive interaction results in sharp resonances in 2D static susceptibilities already in first-order perturbation theory. 
The resonances appear in the form of a logarithmically enhanced square-root non-analyticity near $q = 2 k_F$, where $k_F$ is the Fermi momentum.
This intriguing result shows that 2D semiconductors are strongly susceptible to $2 k_F$ spin- and charge-density fluctuations.

A previous numerical study found that the charge susceptibility has a resonant Kohn anomaly~\cite{kohn1959,kohn1962,afanasev} when evaluated for a screened Coulomb interaction~\cite{ashcroft}. 
However, the situation for the spin susceptibility is far from clear. While the charge and spin susceptibilities are the same for non-interacting systems, it is well known that this is no longer the case in the presence of interactions. In particular, it was shown that the analytic behavior of the two quantities is dramatically different for short-ranged interactions at small $q$ in second-order perturbation theory~\cite{chubmaslov2003}: while the charge suscpetibility stays analytic, the spin susceptibility becomes non-analytic. Surprisingly, this distinction has not been studied for long-range interactions, even at first-order in perturbation theory. The present work aims to fill that gap. We show that the spin susceptibility exhibits a strong Kohn anomaly and features non-analyticities. We provide, for the first time, analytic expressions for logarithmically enhanced interaction corrections to both spin and charge susceptibilities, including the prefactors, which allow us to make predictions for specific materials.
We also show that higher-order interaction corrections to the charge and the spin susceptibilities are different.
In particular, the charge susceptibility is further suppressed by electrostatic screening, while the spin susceptibility is not.
This means $2 k_F$ spin density wave (SDW) ordering is favoured in 2D semiconductors.

There are two main resonant scattering processes near the isotropic FS: forward scattering with small momentum transfer, $q \ll k_F$, and backscattering with momentum transfer, $q \approx 2 k_F$. In the case of a non-interacting isotropic 2DEG, static susceptibilities have a plateau for all momentum transfers $q < 2 k_F$ \cite{afanasev}. Here, we show that the first-order interaction corrections result in sharp resonances in static susceptibilities at the Kohn anomaly that can be further responsible for the DW instabilities of an interacting 2DEG.

Analytic calculations are possible because resonances in static susceptibilities emerge due to scattering processes near the FS that can be accounted for within the semiclassical approximation \cite{lounis,mis2023}: $k_F r \gg 1$, $E_F \tau \gg 1$, where $r$ and $\tau$ are spatial and temporal arguments of a correlation function; $E_F$ is the Fermi energy.
The semiclassical approximation results in a substantial reduction of the spatial degrees of freedom, also known as the dimensional reduction \cite{mis2023}.

The SDW order can be experimentally observed by means of spin-polarized scanning tunneling microscopy (STM) and Raman spectroscopy in a wide range of materials~\cite{heinze,bode,hu2022}.
Even in the absence of a bona fide DW, we expect strong signatures of the enhanced Kohn anomaly. 
The spin susceptibility for a 2DEG in a magnetic field is also exactly equivalent to the layer pseudospin susceptibility in a bilayer 2DEG without a magnetic field. 
Our result then supports the existence of interlayer-dipole-density Friedel oscillations, which can be detected by STM or nitrogen-vacancy centers \cite{torre2016,stano2013}. 
Indeed, Friedel oscillations have been observed in bilayer graphene, silicene, and WSe$_2$~\cite{chen2017,yankowitz2015}. 
These results may also be applicable to transition metal dichalcogenide (TMD) bilayers, as this family of layered van der Waals materials often exhibits DW ordering at low temperatures~\cite{xu2021,chen2016,kidd2002,leroux2015,zhang2022}.
Beyond structures with natural stacking, DW orders have also been observed in twisted bilayers where the DW domain 
is confined to a moir\'e unit cell~\cite{zhao2022}.

The paper is organized as follows. 
The theoretical model is introduced in Sec.~\ref{sec:hopping}.
The free-fermion 2D susceptibility is discussed in Sec.~\ref{sec:loworder}.
The first-order interaction corrections to the 2D susceptibilities are calculated in Sec.~\ref{sec:firstorder}.
Further screening of the charge susceptibility compared to the spin susceptibility is discussed in Sec.~\ref{sec:suscept}.
Comments on the three-dimensional (3D) case are provided in Sec.~\ref{sec:3DEG}.
Discussion of the physical consequences of a strongly peaked Kohn anomaly is presented in Sec.~\ref{sec:insta}.
Conclusions are given in Sec.~\ref{sec:conc}.
Details of derivations are outlined in the appendices.

\section{Theoretical model}
\label{sec:hopping}

In this work we consider a 2D semiconductor with quadratic electron dispersion that is coupled to an external magnetic field, $\bm B$, via the Zeeman coupling. The results apply equally well to the case of hole-doped semiconductors, but for concreteness we consider here positive dispersions.
Such a system is described by the Hamiltonian
\begin{equation}
H_0 = \frac{p^2}{2 m} - E_F + \mbox{\boldmath{$\eta$}} \cdot \bm B ,\label{eq:interhop}
\end{equation}
where $\bm p = (p_x, p_y)$ is the 2D electron momentum, $p = |\bm p|$,
$m$ is the effective mass, $E_F$ is the Fermi energy,
$\mbox{\boldmath{$\eta$}} = \left(\eta_x, \eta_y, \eta_z\right)$ are the spin Pauli matrices, and $\bm B$ is an external magnetic field.
In semiconductor quantum wells $B = |\bm B|$ plays role of the standard Zeeman energy due to the coupling between the external magnetic field and the electron spin.
In general, we choose the $x$ axis directed along the in-plane component of $\bm B$, i.e.
\begin{eqnarray}
&& \bm B = \left(B_x, 0, B_z\right) . \label{Bgen}
\end{eqnarray}
In the case of bilayer materials, the spin corresponds to the layer number such that components of the effective ``magnetic field'' $\bm B$ correspond to the electrostatic bias $\alpha = B_z$ between the layers and the interlayer hopping $t_\perp = B_x$,
\begin{eqnarray}
&& \bm B = \left(t_\perp, 0, \alpha\right) . \label{Bbi}
\end{eqnarray}
Such a model is qualitatively suitable for doped bilayer and heterobilayer TMDs~\cite{liu2013,jones2014,leisgang2020,Zhang_2021,Ghatak:2020aa,C8CP05522J,zhang2017,XIA20171} as well as doped Bernal (AB-stacked) bilayer graphene~\cite{mccann,rozhkov,szabo} at zero magnetic field.
The valley degree of freedom that is typical for the honeycomb lattice materials merely contributes to the degeneracy factor of each band.

In this paper we show that leading resonant contributions to static susceptibilities originate from the forward-scattering interaction with an effective radius $R_0 \gg \lambda_F$, $\lambda_F = 2 \pi /k_F$ is the Fermi wavelength.
In particular, any zero-range interaction such as intervalley and interlayer dipole interactions can be safely neglected here.
Here, we consider any interaction with an effective radius $R_0 \lesssim \lambda_F$ to be zero-ranged.
Finally, we stress that the Zeeman field $\bm{B}$ in Eq.~(\ref{eq:interhop}) can be momentum-dependent, i.e., it may also include spin-orbit interaction.
Indeed, the forward-scattering component of the electron-electron interaction that results in the most resonant contribution to susceptibilities, transfers a small momentum $q \lesssim 1/R_0 \ll k_F$, such that the spinors of scattered electron states are almost collinear and the corresponding spinor matrix elements are always close to one regardless of the symmetry of the spin-orbit interaction.

The single-particle Hamiltonian $H_0$ is diagonal in the $\psi_s$ basis, here $s = \pm 1$ is the band index, 
\begin{equation}
\psi_- = \left(
\begin{array}{c}
\displaystyle \cos\frac{\theta}{2} \\[10pt]
\displaystyle \sin \frac{\theta}{2}
\end{array}
\right) , \hspace{5pt}
\psi_+ = \left(
\begin{array}{c}
\displaystyle \sin\frac{\theta}{2} \\[10pt]
\displaystyle -\cos \frac{\theta}{2}
\end{array}
\right) , \label{psi}
\end{equation}
where we introduced the mixing angle $\theta$,
\begin{equation}
e^{i \theta} = \frac{B_z + i B_x}{B} , \label{theta}
\end{equation}
and $B = |\bm B|$.
Therefore, it is more convenient to work within the following rotated basis made of $\psi_s$ spinors
\begin{eqnarray}
&& \eta_1 = - \eta_x \cos \theta + \eta_z \sin\theta , \label{eta1} \\
&& \eta_2 = \eta_y , \label{eta2} \\
&& \eta_3 = - \eta_x \sin \theta - \eta_z \cos\theta . \label{eta3}
\end{eqnarray}

The spectrum of $H_0$ consists of two parabolic bands,
\begin{equation}
\varepsilon_s(p) = \frac{p^2}{2 m} - E_F - s B , \label{spec}
\end{equation}
where the band index $s = \pm 1$ is also the eigenvalue of $\eta_3$.
The Fermi momenta $k_s$ defined from the two components of the FS via the equation $\varepsilon_s(k_s) = 0$ are then given by
\begin{eqnarray}
&& k_s = k_F + s \, \delta , \label{ks}\\
&& k_F = \frac{k_+ + k_-}{2}, \label{k_F} \\
&& \delta = \frac{k_+ - k_-}{2} = \frac{B}{v_F} , \label{delta}
\end{eqnarray}
where $v_F = k_F/ m$ is the average Fermi velocity, $k_F$ is the average Fermi momentum.
Notice that here we did not define $k_F$ as $\sqrt{2 m E_F}$ but rather introduced it as an average of the two Fermi momenta.
The electron dispersion near the FS is linear in  leading order,
\begin{eqnarray}
&& \varepsilon_s (p) \approx v_s \left(p - k_s \right) , \label{epslin}
\end{eqnarray}
where $v_s = k_s/ m$ is the corresponding Fermi velocity. 
Again, the same diagonalization procedure can be done for an arbitrary momentum-dependent field $\bm B = \bm B_{\bm p}$ at $B_{\bm p} \ll E_F$, where the last condition is required to ensure that two spinors with close (but not necessarily the same) momenta and different band indices are nearly orthogonal.

In this paper we study interaction corrections to static susceptibilities using the Green's function formalism.
The free-fermion Matsubara Green's function $G(\bm r, \tau)$ is diagonal in the $\psi_s$ basis, see Eq.~(\ref{psi}),
\begin{eqnarray}
&& G(\bm r, \tau) = \sum\limits_{s = \pm 1} \psi_s \psi_s^\dagger G_s(r, \tau) , \label{eq:greentot}
\end{eqnarray}
where $G_s(r, \tau)$ is the Green's function corresponding to the band with index $s$, $\tau$ is the imaginary time.
Throughout the paper, we assume the zero temperature limit.
The semiclassical asymptotics of the Green's function at $k_s r \gg 1$ and $E_F \tau \gg 1$ is given by the following expression, see Refs.~\cite{lounis,mis2022,mis2023} for details, 
\begin{eqnarray}
&& G_s (r, \tau) = \sum\limits_{\nu = \pm 1} \frac{e^{i \nu \left(k_s r - \pi/4\right)}}{\sqrt{\lambda_s r}} g_s^\nu(r, \tau) , \label{eq:green} \\
&& g_s^\nu (x, \tau) = \frac{1}{2 \pi} \frac{1}{i \nu x - v_s \tau} , \label{gs}
\end{eqnarray}
where $\lambda_s = 2 \pi / k_s$ and $v_s = k_s/ m$ are the Fermi wavelength and the Fermi velocity corresponding to the band with index $s$, respectively.
The index $\nu$ can be interpreted as the chirality index: $\nu = +1$ ($\nu = -1$) is referred to as the right-handed (left-handed) electron chirality, see Refs.~\cite{mis2022,mis2023}.
Here we also stress that Eqs.~(\ref{eq:green}) and (\ref{gs}) account explicitly for the FS curvature, while the electron dispersion near the FS is linearized according to Eq.~(\ref{epslin}).
The subleading $\propto (p - k_s)^2/m$ correction to $\varepsilon_s(p)$ near $p = k_s$ is negligible in the semiclassical limit $k_F r \gg 1$.

In this work we consider a repulsive instantaneous density-density interaction between electrons
\begin{eqnarray}
&& V(r, \tau) = V(r) \delta(\tau) , \label{Vdelta}
\end{eqnarray}
where $\delta(\tau)$ is the delta function, $V(r)$ is a function of $r$.
We are specifically interested in two cases. The first is the Coulomb interaction 
\begin{eqnarray}
&& V_C (r) = \frac{e^2}{\epsilon r} , \label{Coulomb}
\end{eqnarray}
where $e$ is the elementary charge, and $\epsilon$ is the dielectric constant. The second is an arbitrary finite-range interaction, $V_{R_0}(r)$, where $R_0 \gg \lambda_F$ denotes an effective radius of the interaction, $\lambda_F = 2 \pi/k_F$ is the Fermi wavelength. 
As a particular example of the latter, we use the screened Thomas-Fermi interaction in 2D, $V_{TF}(r)$,
\begin{eqnarray}
&& V_{TF} (r) = \int\limits_0^\infty\frac{dq}{2\pi} \, q J_0(q r) V_{TF}(q) , \label{VTFr} \\
&& V_{TF}(q) = \frac{2 \pi e^2}{\epsilon} \frac{1}{q + \kappa} ,\label{VTFq} \\
&&  \kappa = \frac{2 \pi e^2 N_F}{\epsilon} = \frac{2\mathfrak{g}}{a_B} , \hspace{5pt} a_B = \frac{\epsilon}{m e^2} , \label{kappa} \\
&& N_F = \frac{\mathfrak{g} m}{\pi} , \label{NF}
\end{eqnarray}
where Eq.~(\ref{VTFr}) is the 2D Fourier transform, $J_0(x)$ is a Bessel function, $1/\kappa = R_0 \gg \lambda_F$ defines the effective radius of the Thomas-Fermi interaction, $a_B$ is the effective Bohr radius, $\mathfrak{g}$ is the degeneracy of each FS, and $N_F$ is the total density of states at the Fermi energy.

Note that the effective Bohr radius $a_B$ is a few nanometers for TMD bilayers, Silicon MOSFETs and GeS monolayers~\cite{padhi2021, zhang2005, luo2020}. Several monochalcogenides as well as  bilayer graphene with a Boron Nitride spacer have effective Bohr radii of 5-9nm~\cite{luo2020, chui2020}, while quantum wells typically do better with $a_B=8$nm, $10$nm, and $18$nm for GaAs, AlGaAs, and strained Ge/Si  wells, respectively~\cite{zhang2005, bokes2010, conti2021,scappucci2021}. Ideal candidates are 2DEGs surrounded by a high-dielectric environment to fulfill the condition $a_B\gg\lambda_F$. Indeed some semiconductors have dielectric constants over 1000~\cite{osada2012}. One would need to tune the density via gates or chemical doping into this regime to see the predicted susceptibility that will be derived in Sec.~\ref{sec:q2kf}.

It is convenient to introduce the Wigner-Seitz radius $r_s$, which plays the role of a dimensionless coupling constant for the interactions,
\begin{eqnarray}
&& r_s \equiv \frac{\sqrt{\mathfrak{g}}}{a_B \sqrt{\pi n_e}} \approx \frac{\sqrt{2}}{k_F a_B} , \label{rs}
\end{eqnarray}
where $n_e$ is the 2D electron density. In the approximate relation we implied $\delta \ll k_F$, i.e., the Fermi surfaces are only slightly split by the effective Zeeman term, so $n_e \approx \mathfrak{g} k_F^2/(2 \pi)$.
We also point out that screening of the Coulomb interaction in real devices has contributions from the proximity to metallic gates that control the electron density, such that $R_0$ in real devices is bound from above by the distance to the closest metallic gate.
The Thomas-Fermi screening does not account for dynamic screening effects. 
Such effects can be taken into account within second order perturbation theory, see, e.g., Ref.~\cite{zak2012}.
As here we are interested in the first-order interaction corrections to the susceptibilities, we can neglect dynamic screening as long as $r_s \ll 1$. Note that this perturbative regime is consistent with the condition $a_B\gg\lambda_F$ discussed earlier.

\begin{figure}
	\includegraphics[width=0.8\columnwidth]{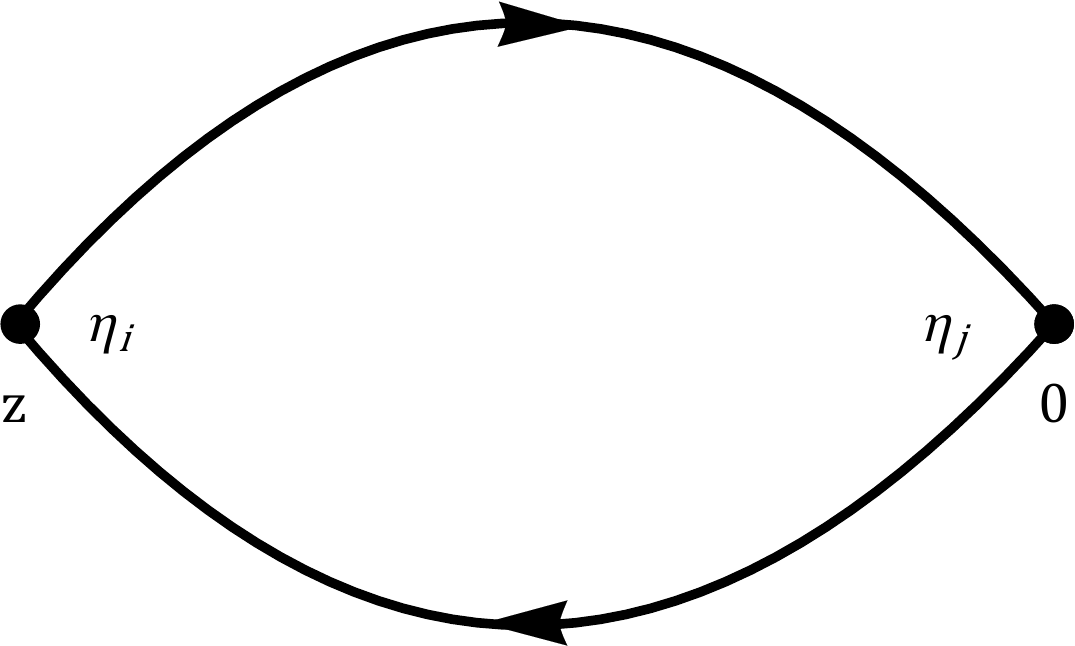}
	\caption{Free-fermion susceptibility $\chi_{ij}^{(0)}(z)$, $z = (\bm r, \tau)$ [see Eqs.~(\ref{Pi0}) and (\ref{chissprime0})]. Here, $\eta_i$ is either a Pauli matrix for $i \in \{1,2,3\}$ [see Eqs.~(\ref{eta1})--(\ref{eta3})] or the identity matrix, $\eta_0 = I$; the same applies to $\eta_j$.}
	\label{fig:free}
\end{figure}

\section{Free-fermion susceptibility}
\label{sec:loworder}

The free-fermion susceptibility is represented by the bubble diagram, see Fig.~\ref{fig:free},
\begin{eqnarray}
&& \chi_{ij}^{(0)}(r, \tau) = -{\rm Tr} \left\{\eta_i G(r, \tau) \eta_j G(r, -\tau) \right\} \nonumber \\
&& \hspace{42pt} = \sum\limits_{s,s'} \eta_i^{ss'} \eta_j^{s's} \chi_{s's}^{(0)}(r, \tau) , \label{Pi0} \\
&& \chi_{s's}^{(0)}(r, \tau) = - \mathfrak{g} \, G_{s'}(r, \tau) G_s(r, -\tau) , \label{chissprime0}
\end{eqnarray}
where $G_s(r, \tau)$ is defined in Eq.~(\ref{eq:greentot}), its semiclassical asymptotics is given by Eq.~(\ref{eq:green}), 
$\eta_i^{ss'} = \langle \psi_s|\eta_i |\psi_{s'}\rangle$, $i, j \in \{0,1,2,3\}$, where $\eta_i$ is a Pauli matrix if $i \in \{1,2,3\}$, see Eqs.~(\ref{eta1})--(\ref{eta3}), and $\eta_0 = I$, where $I$ is the $2 \times 2$ identity matrix.
${\rm Tr}$ stands for the trace over the spin as well as any other possible degree of freedom; the latter ones are accounted for by the degeneracy factor $\mathfrak{g}$ of each FS. $\mathfrak{g} = 2$ in Si MOSFETs due to the valley degeneracy, $\mathfrak{g} = 1$ in III-V semiconducting quantum wells. Note that for the case of the layer pseudospin susceptibility in a bilayer TMD, $\mathfrak{g} = 4$ due to the electron spin and valley degeneracies (in this case we treat $\mbox{\boldmath{$\eta$}}$ as the layer pseudospin).

The Fourier transform of Eq.~(\ref{chissprime0}) at zero frequency yields the 2D static susceptibility, see Ref.~\cite{afanasev} and Appendix~\ref{app:free}:
\begin{eqnarray}
&&\hspace{-15pt} \chi_{s's}^{(0)}(q) = \frac{N_F}{2} \nonumber \\
&& \hspace{-5pt} \times \left[1 - \vartheta \left(\delta q\right) \sqrt{\left(1 - \frac{[Q_{s's}^{-}]^2}{q^2}\right) \left(1 - \frac{[Q_{s's}^{+}]^2}{q^2}\right)}\right] , \label{chiq0} \\
&& \hspace{-15pt} Q_{s's}^\pm = |k_{s'} \pm k_s| , \label{Qpm}
\end{eqnarray}
where $\vartheta(x)$ is the Heaviside step function, $\delta q = q - Q_{s's}^+$, $N_F$ is the density of states, see Eq.~(\ref{NF}).

\begin{figure}
	\includegraphics[width=0.45\columnwidth]{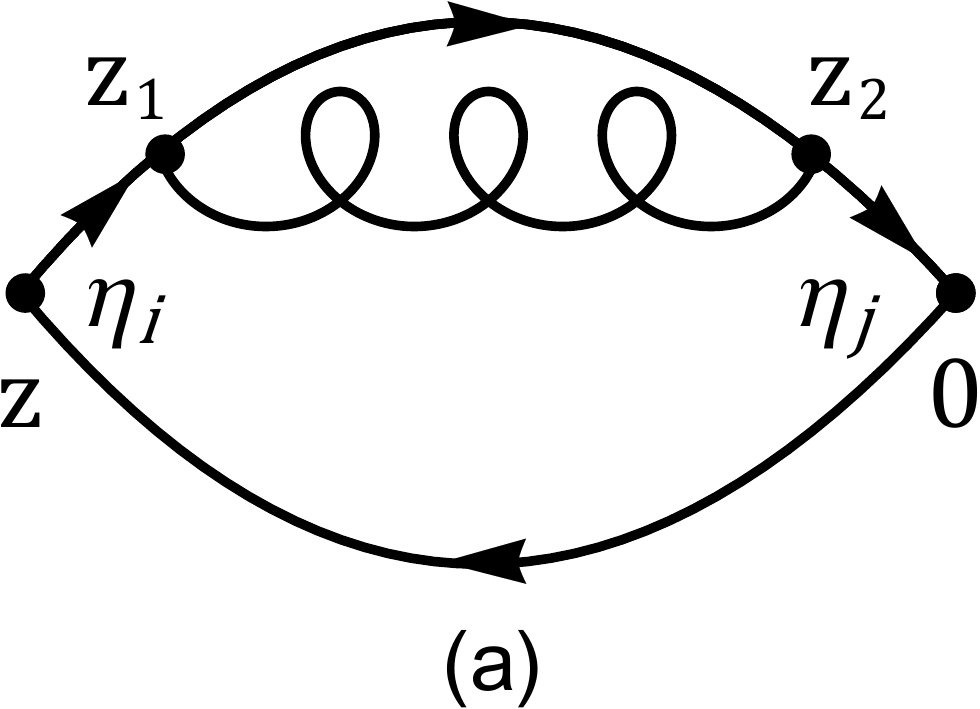} \hspace{5pt} \includegraphics[width=0.45\columnwidth]{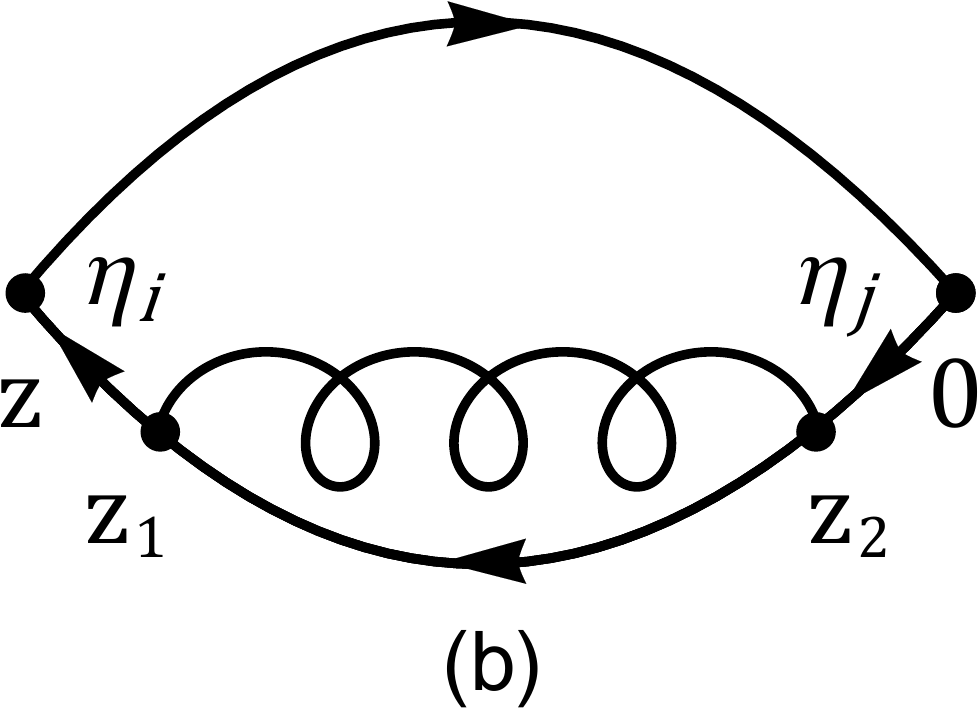}\\
	\includegraphics[width=0.45\columnwidth]{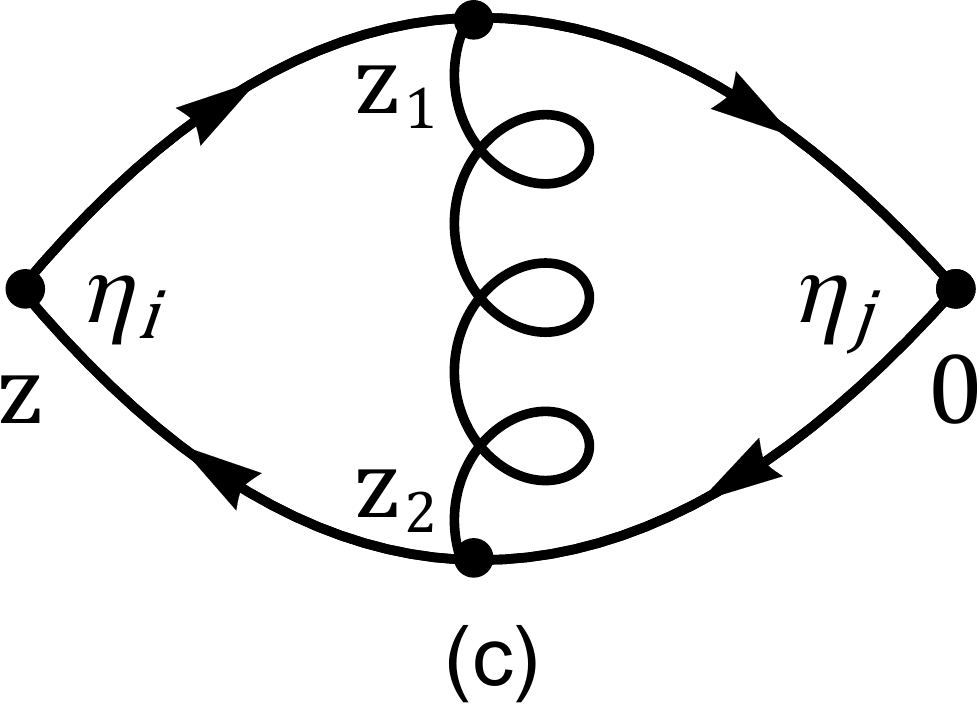}
	\caption{First-order diagrams for the irreducible susceptibility, $\chi_{ij}^\ell(z)$, $z = (\bm r, \tau)$, $\ell \in \{a, b, c\}$, see Eqs.~(\ref{chia}), (\ref{chib}), and (\ref{chic}).}
	\label{fig:firstorder}
\end{figure}

The free-fermion susceptibility has a one-sided square-root non-analyticity at $q > Q_{s's}^+$ that does not provide a peak structure: the maximal value of $\chi_{s's}^{(0)}(q)$ is given by $N_F/2$ at all $q < Q_{s's}^+$, where $Q_{s's}^+ \approx 2 k_F$. This is the usual Lindhard function.
Thus, the interaction corrections determine the resonant value of $q$ where the susceptibility reaches its maximum.
This is especially important for studying possible DW instabilities of the 2DEG.

In the next sections we show that the first-order interaction correction to the layer susceptibility yields a two-sided logarithmically enhanced square-root non-analyticity with an asymmetric peak structure near the Kohn anomaly $q = Q_{s's}^+$, even for a finite range interaction.
Our results agree with a numerical study of the charge susceptibility of spin-degenerate 2DEGs with a dynamically screened Coulomb interaction~\cite{ashcroft}.
We also find a weak non-analyticity at $q = Q_{-ss}^- = 2 \delta$, however, this non-analyticity is removed by the Thomas-Fermi screening of the Coulomb interaction.
The non-analyticity at $q \sim 0$ is canceled  for arbitrary density-density interactions within first-order perturbation theory.
A weak linear-in-$q$ non-analyticity at $q \sim 0$ is possible within the second-order perturbation theory, see, e.g., Refs.~\cite{belitzkirk1997,chubmaslov2003,chubglazman2005,chesizak,zak2012}.
However, the much stronger square-root non-analyticity near the Kohn anomaly that emerges already within the first-order perturbation theory is more important.

\section{First-order interaction corrections}
\label{sec:firstorder}

In this section we calculate the first-order corrections to irreducible susceptibilities.
Irreducible susceptibilities are represented by the sum of all diagrams that remain connected after cutting an arbitrary interaction line.

The first-order interaction corrections to the irreducible susceptibility are given by the three diagrams in Fig.~\ref{fig:firstorder}
\vspace{-2pt}
\begin{eqnarray}
&& \hspace{-20pt} \chi^{a}_{ij}(z) = \int dz_1 dz_2 \, V(z_1 - z_2) \nonumber \\
&& \hspace{9pt} \times {\rm Tr}\left\{\eta_i G(z - z_1) G(z_1 - z_2) G(z_2) \eta_j G(-z)
\right\} , \label{chia} 
\end{eqnarray}
\begin{eqnarray}
&& \hspace{-20pt} \chi^{b}_{ij}(z) = \int dz_1 dz_2 \, V(z_1 - z_2) \nonumber \\
&& \hspace{9pt} \times {\rm Tr}\left\{\eta_i G(z) \eta_j G(- z_2) G(z_2 - z_1) G(z_1 - z)
\right\} , \label{chib}
\end{eqnarray}
\begin{eqnarray}
&& \hspace{-20pt} \chi^{c}_{ij}(z) = \int dz_1 dz_2 \, V(z_1 - z_2) \nonumber \\
&& \hspace{9pt} \times {\rm Tr}\left\{\eta_i G(z - z_1) G(z_1) \eta_j G(-z_2) G(z_2 - z)
\right\} , \label{chic}
\end{eqnarray}
where $z_i = (\bm r_i, \tau_i)$, $V(z) = V(\bm r, \tau) = V(r, \tau)$ is an arbitrary density-density interaction. 
The trace, ${\rm Tr}$, in Eqs.~(\ref{chia})--(\ref{chic}) is evaluated the same way as in Eq.~(\ref{Pi0}), yielding
\begin{eqnarray}
&& \chi_{ij}^\ell(z) = \sum\limits_{s,s'} \eta_i^{ss'} \eta_j^{s's} \chi_{s's}^\ell (z) , \label{chissprime1}
\end{eqnarray}
where $z = (\bm r, \tau)$, $\ell \in \{a,b,c\}$, $\eta_i^{ss'} = \langle \psi_s|\eta_i|\psi_{s'}\rangle$,
and $\chi_{s's}^\ell(z)$ are given by the following integrals:
\begin{eqnarray}
&& \hspace{-12pt} \chi_{s's}^{a}(z) = \chi_{ss'}^b(-z) = \mathfrak{g} \, G_s(-z) \int dz_1 dz_2 \, V(z_1 - z_2) \nonumber \\
&& \hspace{21pt} \times G_{s'}(z - z_1) G_{s'}(z_1 - z_2) G_{s'}(z_2) , \label{chiabssprime} \\
&& \hspace{-12pt} \chi_{s's}^c (z) = \mathfrak{g} \int dz_1 dz_2 \, V(z_1 - z_2) G_{s'}(z - z_1) \nonumber \\
&& \hspace{21pt} \times G_{s'}(z_1) G_{s}(-z_2) G_{s}(z_2 - z) , \label{chicssprime}
\end{eqnarray}
where $\mathfrak{g}$ is the degeneracy factor.

Non-analytic corrections to the susceptibilities originate from resonant scattering processes near the FS. 
The leading non-analytic contribution of each Feynman diagram can be evaluated exactly within the semiclassical approximation $k_F r \gg 1$, $E_F \tau \gg 1$.
Using the semiclassical asymptotics of the Green's function, see Eq.~(\ref{eq:green}), we perform the dimensional reduction of the diagrams in Fig.~\ref{fig:firstorder} following the recipe described in Ref.~\cite{mis2023}. The details are given in Appendix~\ref{app:dimred} and we find
\begin{eqnarray}
&& \hspace{-7pt} \chi^\ell_{s's}(r, \tau) = \mathfrak{g} \sum\limits_{\nu, \nu'} \frac{e^{i \frac{\pi}{4}(\nu -\nu')}}{\sqrt{\lambda_s \lambda_{s'}}} \frac{e^{i r Q_{\gamma'\gamma}}}{r} \tilde \chi^\ell_{\gamma' \gamma}(r, \tau) , \label{chired} \\
&& \hspace{-7pt} Q_{\gamma'\gamma} \equiv \nu' k_{s'} - \nu k_{s} = \left(\nu' - \nu\right) k_F + \left(\nu' s' - \nu s\right) \delta. \label{Q}
\end{eqnarray}
Here, the collective indices $\gamma = \{\nu, s\}$, $\gamma' = \{\nu', s'\}$ are introduced to shorten notations, $\lambda_s = 2 \pi/ k_s$, $k_s$ is the Fermi momentum of the $s^{\rm th}$ FS, see Eq.~(\ref{ks}), $\nu, \nu' \in \{\pm 1\}$ are the chiral indices [see the discussion after Eq.~(\ref{gs})], $\ell \in \{a, b, c\}$, and $\tilde \chi^\ell_{\gamma'\gamma}(x, \tau)$ are represented by the one-dimensional analogues of the diagrams in Fig.~\ref{fig:firstorder}
\begin{eqnarray}
&& \hspace{-12pt} \tilde \chi^a_{\gamma'\gamma}(\xi) = \tilde \chi^b_{\gamma \gamma'}(-\xi) = g_\gamma(-\xi) \int d\xi_1 d\xi_2 \, V(\xi_1 - \xi_2)  \nonumber \\
&& \hspace{22pt} \times g_{\gamma'} (\xi - \xi_1) g_{\gamma'}(\xi_1 - \xi_2) g_{\gamma'}(\xi_2) , \label{chiab1D} \\
&& \hspace{-12pt} \tilde \chi^c_{\gamma'\gamma}(\xi) = \int d\xi_1 d\xi_2 \, V(\xi_1 - \xi_2) g_{\gamma'}(\xi - \xi_1) \nonumber \\
&& \hspace{22pt} \times g_{\gamma'}(\xi_1) g_{\gamma}(-\xi_2) g_{\gamma} (\xi_2 - \xi) , \label{chic1D}
\end{eqnarray}
where $\xi_i = (x_i, \tau_i)$, and $d\xi_i = dx_i d\tau_i$. Here, $x_i \in (-\infty, \infty) = \mathbb{R}$ is an effective one-dimensional coordinate so that integrals over $x_1$ and $x_2$ are taken over $\mathbb{R}$.
We have also introduced a short-hand notation for the one-dimensional Green's function and the interaction,
\begin{eqnarray}
&& g_\gamma (\xi) = g_s^\nu(x, \tau) , \label{ggamma} \\
&& V(\xi) = V(|x|, \tau) , \label{Vx}
\end{eqnarray}
where $g_s^\nu(x, \tau)$ is given by Eq.~(\ref{gs}).
The derivation of Eqs.~(\ref{chiab1D}) and (\ref{chic1D}) is outlined in Appendix~\ref{app:dimred}.
Thus, the dimensional reduction technique (see Ref.~\cite{mis2023}), allowed us to reduce the first-order diagrams in Fig.~\ref{fig:firstorder} to their one-dimensional analogs given by Eqs.~(\ref{chiab1D}) and (\ref{chic1D}).
The 2D susceptibilities are then expressed through the one-dimensional ones via Eq.~(\ref{chired}).

Now, we proceed to calculate the static susceptibilities in the zero-temperature limit for instantaneous interactions, see Eq.~(\ref{Vdelta}). The one-dimensional static susceptibilities $\tilde \chi^\ell (Q_{\gamma' \gamma}, x)$ are given by the following expression:
\begin{eqnarray}
&& \tilde \chi^\ell \left(Q_{\gamma' \gamma}, x\right) \equiv \int\limits_{-\infty}^\infty \tilde \chi^\ell_{\gamma' \gamma} (x, \tau) \, d\tau . \label{static1D}
\end{eqnarray}
The non-analytic contribution to the 2D static susceptibilities then follows directly from Eq.~(\ref{chired}),
\begin{eqnarray}
&& \hspace{-25pt} \chi^\ell_{s's}(r) \equiv \int\limits_{-\infty}^\infty \chi^\ell_{s' s} (r, \tau) \, d\tau \nonumber \\
&& \hspace{7pt} = \mathfrak{g} \sum\limits_{\nu, \nu'} \frac{e^{i \frac{\pi}{4}(\nu -\nu')}}{\sqrt{\lambda_s \lambda_{s'}}} \frac{e^{i r Q_{\gamma'\gamma}}}{r} \tilde \chi^\ell\left(Q_{\gamma' \gamma}, r\right) , \label{static2D}
\end{eqnarray}
where $\tilde \chi^\ell\left(Q_{\gamma' \gamma}, r\right)$ is given by Eq.~(\ref{static1D}) at $x = r$ and represents the amplitude of the $Q_{\gamma' \gamma}$ harmonic of the 2D static susceptibility.  
All time integrals can be evaluated analytically for an arbitrary instantaneous interaction [Eq.~(\ref{Vdelta})], see Appendix~\ref{app:1D} for details,
\begin{eqnarray}
&& \hspace{-10pt} \tilde \chi^a\left(Q_{\gamma' \gamma}, x\right) = \tilde \chi^b\left(Q_{\gamma \gamma'}, x\right) = \int \frac{d x_1 dx_2}{4 \pi^2} \frac{V(x_1 - x_2)}{v_{s'} \left(x_1 - x_2\right)} \nonumber \\
&& \hspace{25pt} \times \frac{\vartheta\left(x_2\right) \vartheta\left(-\nu \nu' |x| - x_1\right)}{v_{s'} |x| + v_s \left(|x_2| + \left|x_1 + \nu \nu' |x|\right|\right)} , \label{chiabstatic1D} \\
&& \hspace{-10pt} \tilde \chi^c\left(Q_{\gamma' \gamma}, x\right) = \int \frac{dx_1 dx_2}{4 \pi^2} \frac{V(x_1 - x_2)}{v_s |x_1| + v_{s'} |x_2|} \nonumber \\
&& \hspace{25pt} \times \frac{\vartheta\left(- \nu \nu' x_1 x_2\right)\vartheta\left(- \nu \nu' \left(x - x_1\right) \left(x - x_2\right)\right)}{v_s \left|x - x_1\right| + v_{s'} \left|x - x_2\right|} , \label{chicstatic1D}
\end{eqnarray}
where $\vartheta(x)$ is the Heaviside step function, $v_s = k_s/m$ is the Fermi velocity at the $s^{\rm th}$ FS, and integrals over $x_1$ and $x_2$ are taken over the real line $\mathbb{R}$.
At this point it is more convenient to work with the sum of all three first-order contributions,
\begin{eqnarray}
&& \tilde \chi^{(1)} \left(Q_{\gamma' \gamma}, x\right) = \sum\limits_{\ell \in \{a,b,c\}} \tilde \chi^{\ell} \left(Q_{\gamma' \gamma}, x\right) , \label{chisum}
\end{eqnarray}
where the superscript $^{(1)}$ indicates the first order correction due to interactions.
The remaining integrals over $x_1$ and $x_2$ in Eqs.~(\ref{chiabstatic1D}), (\ref{chicstatic1D}) can be further simplified for arbitrary $V(x)$, see Appendix~\ref{app:1D} for details,
\begin{eqnarray}
&& \hspace{-13pt} \tilde \chi^{(1)} \left(Q_{\gamma' \gamma}, r\right) = \frac{1}{4 \pi^2 v_s v_{s'} r} \int\limits_{-\sigma r}^{+\infty} dx \, V(x) \mathcal{F}_{s's}^{\sigma} \left(\frac{x}{r}\right) , \label{chisumr} \\
&& \hspace{-13pt} \mathcal{F}^\sigma_{s's}(y) = \frac{2}{a_\sigma} \ln\left|1 + a_\sigma \frac{y + \sigma}{y^2}\right| - \frac{\left(y + \sigma\right) \left(2 y + a_\sigma \right)}{y \left[y^2 + a_\sigma (y + \sigma) \right]} , \label{F} \\
&& \hspace{-13pt} a_\sigma = \frac{\left(k_s + \sigma k_{s'}\right)^2}{k_s k_{s'}} , \label{a}
\end{eqnarray}
where $\sigma = - \nu \nu'$, $y \in \mathbb{R}$. 
Next, we discuss the anomalies at $q \sim 0$, $q = 2 \delta$ and $q = k_s + k_{s'}$ separately.

\subsection{No anomaly at $q \sim 0$}

Here we consider the contribution to the susceptibilities at $q \sim 0$ that correspond to $Q_{\gamma' \gamma} = 0$ [see Eq.~(\ref{Q})].
This condition is met if $\nu' = \nu$ and $s' = s$ which results in $\sigma = -1$ and $a_\sigma = 0$ in Eqs.~(\ref{chisumr})--(\ref{a}).
Taking the limit $a_\sigma \to 0$ in Eq.~(\ref{F}), we find that $\mathcal{F}_{s's}^\sigma(y) = \mathcal{F}_{ss}^-(y) = 0$ under these conditions.
Therefore, there is no non-analyticity/discontinuity of the first-order interaction contribution to the susceptibilities at $q = 0$.
Here we stress that the exact value of the first-order interaction correction to static susceptibilities at $q = 0$ is not necessarily zero due to the contributions of the virtual scattering processes far away from the FS, however, those contributions must be analytic in the vicinity of $q = 0$, i.e. there is no resonant structure of the susceptibilities there.
Non-analytic contributions to the static susceptibilities at $q \sim 0$ may appear from the dynamic screening of the interaction; this can be accounted for via second-order perturbation theory and results in small linear-in-$q$ correction, see, e.g. Refs.~\cite{belitzkirk1997,chubmaslov2003,chubglazman2005,chesizak,zak2012}.

\subsection{Weak Kohn anomaly at $q = 2 \delta$}

Next, we consider $|Q_{\gamma'\gamma}| = 2 \delta$ [see Eq.~(\ref{Q})].
This condition is met when $\nu' = \nu$ and $s' = - s$, which corresponds to $\sigma = - 1$ and $a_\sigma = 4 \delta^2/(k_+ k_-)$ in Eqs.~(\ref{chisumr}), (\ref{F}).
If the Zeeman splitting of the FS is small, $\delta \ll k_F$, then $a_\sigma \ll 1$ and $\mathcal{F}^\sigma_{s's}(y)$ can be expanded in powers of $a_\sigma$. The leading term is the following
\begin{eqnarray}
&& \mathcal{F}^-_{-ss}(y) \approx - \frac{4 \delta^2}{k_F^2} \frac{y - 1}{y^4} , \label{Fminus}
\end{eqnarray}
where $\delta \ll k_F$.
Substituting Eq.~(\ref{Fminus}) into Eq.~(\ref{chisumr}), we find the $2 \delta$ harmonic $\tilde \chi^{(1)} (\pm 2 \delta, r) = \tilde \chi^{(1)} (2 \delta, r)$,
\begin{eqnarray}
&& \hspace{-5pt} \tilde \chi^{(1)} (2 \delta, r) \approx - \left(\frac{\delta}{\pi v_F k_F}\right)^2 \int\limits_1^{+\infty} dy \, V(r y) \frac{y - 1}{y^4} , \label{chiminus}
\end{eqnarray} 
where $Q_{\gamma'\gamma} = \pm 2 \delta$ at $\nu' = \nu$ and $s' = - s$.
We also introduced a new integration variable $y = x/r$ in Eq.~(\ref{chisumr}).

First, we consider the Coulomb interaction [see Eq.~(\ref{Coulomb})] which we insert into Eq.~(\ref{chiminus}) to get
\begin{eqnarray}
&& \tilde \chi^{(1)} (2 \delta, r) \approx - \frac{\sqrt{2} r_s}{24 \pi^2 v_F} \left(\frac{\delta}{k_F}\right)^2 \frac{1}{r} , \label{chiminuscoulomb}
\end{eqnarray}
where $r_s$ is defined in Eq.~(\ref{rs}).
Summing over all three first-order diagrams in Eq.~(\ref{static2D}) and using Eq.~(\ref{chiminuscoulomb}), we find the $2 \delta$ harmonic, $\chi_{-ss}^{(1)}(2 \delta, r)$, in the large-distance asymptotics of the static 2D susceptibility that corresponds to $\nu' = \nu$ and $s' = -s$ in Eq.~(\ref{static2D}), given by
\begin{eqnarray}
&& \hspace{-30pt} \chi_{-ss}^{(1)}(2 \delta, r) \approx \frac{\mathfrak{g}}{\lambda_F} \sum\limits_\nu \frac{e^{-2 i \nu s \delta \, r}}{r} \tilde \chi^{(1)}(2 \delta, r) \nonumber \\
&& \hspace{20pt} \approx - \frac{\sqrt{2} r_s N_F}{24 \pi^2} \left(\frac{\delta}{k_F}\right)^2 \frac{\cos\left(2 \delta r\right)}{r^2} , \label{2Dminusr} 
\end{eqnarray}
where we used $\lambda_+ \approx \lambda_- \approx \lambda_F = 2\pi/k_F$. $N_F$ is the total density of states at the Fermi energy [see Eq.~(\ref{NF})].
Indeed, we find a weak Kohn anomaly in Eq.~(\ref{2Dminusr}).
The Fourier transform of Eq.~(\ref{2Dminusr}) (see Supplementary Information Section~4) results in a weak one-sided square-root non-analyticity at $q = 2\delta$,
\begin{eqnarray}
&& \hspace{-15pt} \chi_{-ss}^{(1)}(2 \delta, q) \approx \frac{r_s N_F}{6 \pi} \left(\frac{\delta}{k_F}\right)^2 \vartheta\left(\frac{2 \delta}{q} - 1\right) \sqrt{\frac{2 \delta - q}{2 \delta}} , \label{2Dminusq}
\end{eqnarray}
where $q - 2 \delta \ll \delta$ and $\vartheta(x)$ is the Heaviside step function.
This non-analyticity does not produce a resonant peak structure near $q = 2 \delta$ due to its positive sign and the one-sided character.
Moreover, this non-analyticity is strongly suppressed by the factor $(\delta/k_F)^2 \ll 1$.

The screened 2D interaction in Eqs.~(\ref{VTFr}), (\ref{VTFq}), has asymptotics $V_{TF}(r) \propto 1/r^3$ for $r \gg 1/\kappa$, where $\kappa$ is defined in Eq.~(\ref{kappa}). 
From this, one can easily verify that the non-analyticity produced is extremely weak: $\chi_{-ss}^{(1)}(2 \delta, q) \propto |2 \delta - q|^{5/2}$ at $|q - 2 \delta| \ll {\rm min}(\delta, \kappa)$, which does not result in a resonant peak structure near $q = 2 \delta$.

\subsection{Resonant Kohn anomaly at $q \approx 2 k_F$}\label{sec:q2kf}

Finally, we consider the case when $|Q_{\gamma' \gamma}| = k_s + k_{s'} \approx 2 k_F$, see Eq.~(\ref{Q}), which corresponds to $\nu' = - \nu$ for arbitrary $s$ and $s'$.
In this case, $\sigma = +1$ in Eq.~(\ref{chisumr}) and $a_\sigma \approx 4$ for $\delta \ll k_F$.
This allows us to further simplify the function $\mathcal{F}^\sigma_{s's}(y)$,
\begin{eqnarray}
&& \mathcal{F}^+_{s's}(y) \approx \ln \left|1 + \frac{2}{y}\right| - \frac{2 (y + 1)}{y (y + 2)} \equiv \mathcal{F}^+(y) , \label{Fplus}
\end{eqnarray}
where we used $a_+ \approx 4$ in Eqs.~(\ref{F}), (\ref{a}).
Here we introduced the new notation $\mathcal{F}^+(y)$, since the dependence on $s$ and $s'$ is negligible in the limit $\delta \ll k_F$.
Substituting Eq.~(\ref{chisumr}) into Eq.~(\ref{static2D}) and taking $\nu' = -\nu$, we find the contribution to the static susceptibilities near $q = Q_{s's}^+ = k_{s} + k_{s'}$, where $Q_{s's}^+$ is defined in Eq.~(\ref{Qpm}),
\begin{eqnarray}
&& \chi_{s's}^{(1)}\left(Q_{s's}^+, r\right) \approx \frac{\mathfrak{g}}{\lambda_F} \sum\limits_\nu \frac{e^{i \nu \left(\frac{\pi}{2} - Q_{s's}^+ r\right)}}{r} \tilde \chi^{(1)}\left(Q_{s's}^+, r\right) \nonumber \\
&& = \frac{N_F}{4 \pi^2 v_F} \frac{\sin \left(Q_{s's}^+ r\right)}{r} \int\limits_{-1}^{+\infty} dy \, V(r y) \mathcal{F}^+(y) . \label{chi2kFr}
\end{eqnarray}
Here, $\tilde \chi^{(1)}(Q_{s's}^+, r)$ is given by Eq.~(\ref{chisumr}) at $\sigma = +1$ and $|Q_{\gamma'\gamma}| = Q_{s's}^+$. Furthermore, we used $\lambda_+ \approx \lambda_- \approx \lambda_F = 2\pi/k_F$ and $Q_{\gamma'\gamma} = -\nu Q_{s's}^+$ at $\nu' = -\nu$.

First, we study Eq.~(\ref{chi2kFr}) for the Coulomb interaction given by Eq.~(\ref{Coulomb}).
Notice that one has to use $V(x) = V(|x|)$ as the interaction depends only on the absolute value of the distance.
It is then clear that the integral over $y$ in Eq.~(\ref{chi2kFr}) is divergent at $y = 0$.
This divergence is unphysical and it originates from the semiclassical expansion $r \gg \lambda_F$ that we used here.
This means that the argument of $V(r y)$ in Eq.~(\ref{chi2kFr}) cannot be too small as the semiclassical approximation is no longer valid at $r |y| \lesssim \lambda_F$, which gives the bound $|y| \gg \lambda_F/r$.
For this reason we introduce the cut-off momentum $k_\Lambda \sim k_F$ such that the integration over $y$ in Eq.~(\ref{chi2kFr}) is taken over the union of two intervals $y \in (-1, - 1/(k_\Lambda r)) \cup (1/(k_\Lambda r), +\infty)$ where $k_\Lambda r \gg 1$.
Substituting the Coulomb interaction into Eq.~(\ref{chi2kFr}), we find
\begin{eqnarray}
&& \chi_{s's}^{(1)} \left(Q_{s's}^+, r\right) \approx \frac{\sqrt{2} r_s N_F}{8 \pi^2} \frac{\sin \left(Q_{s's}^+ r\right)}{r^2} \mathcal{I}\left(k_\Lambda r\right) , \label{chiI} \\
&& \mathcal{I}(x) = \int\limits_{-1}^{-1/x} \frac{dy}{|y|} \mathcal{F}^+(y) + \int\limits_{1/y}^{+\infty} \frac{dy}{y} \mathcal{F}^+(y) , \label{I}
\end{eqnarray}
where $\mathcal{F}^+(y)$ is given by Eq.~(\ref{Fplus}) and $r_s$ was introduced in Eq.~(\ref{rs}).
We are interested in the $x \gg 1$ asymptotics of $\mathcal{I}(x)$, 
\begin{eqnarray} 
&& \mathcal{I}(x) = (\ln x)^2 + \left(2 \ln 2 - 1\right) \ln x + \mathcal{O}(1) .
\end{eqnarray}
The leading $\mathcal{I}(x) \approx (\ln x)^2$ asymptotics originates from the vertex correction diagram shown in Fig.~\ref{fig:firstorder}(c).
The diagrams in Figs.~\ref{fig:firstorder}(a) and (b) contribute to the subleading $\propto \ln(x)$ correction.
Within the leading logarithmic order, we find the $Q_{s's}^+$ harmonic of the static susceptibilities
\begin{eqnarray}
&& \chi_{s's}^{(1)} \left(Q_{s's}^+, r\right) \approx \frac{\sqrt{2} r_s N_F}{8 \pi^2} \frac{\sin \left(Q_{s's}^+ r\right)}{r^2} \left[\ln\left(k_\Lambda r\right)\right]^2 , \label{chi2kFrlead}
\end{eqnarray}
where $k_\Lambda \sim k_F$ is the ultraviolet cut-off determining the applicability range of the semiclassical approximation.
The Fourier transform of Eq.~(\ref{chi2kFrlead}) (see Appendix~\ref{app:Fourier}), shows the asymmetric resonant non-analytic peak near $q = Q_{s's}^+ = k_s + k_{s'}$,
\begin{eqnarray}
&& \hspace{-30pt} \chi_{s's}^{(1)}\left(Q_{s's}^+, q\right) \approx - \frac{r_s N_F}{2 \pi} \sqrt{\left|\frac{\delta q}{Q_{s's}^+}\right|} \ln\left|\frac{k_\Lambda}{\delta q}\right| \nonumber \\
&& \hspace{27pt} \times \left[\vartheta\left(\delta q\right) \ln \left|\frac{k_\Lambda}{\delta q}\right|+ \pi \vartheta\left(-\delta q\right)\right] , \label{chi2kFq}
\end{eqnarray}
where $\vartheta(x)$ is the Heaviside step function, $\delta q = q - Q^+_{s's} \ll Q_{s's}^+ \approx 2 k_F$ and $k_\Lambda \sim k_F$.
The peak structure originates from the negative two-sided square-root non-analyticity.
Here we also see that this non-analyticity is enhanced by $(\ln|k_\Lambda/\delta q|)^2$ at $\delta q > 0$ and by $\ln|k_\Lambda/\delta q|$ at $\delta q < 0$, where $\delta q = q - Q_{s's}^+$.

In real devices the Coulomb interaction is screened due to the proximity of the 2DEG to metallic gates, as well as the Thomas-Fermi screening effect.
To account for this, we now consider a general finite-range interaction, $V_{R_0}(r)$, where $R_0 \gg \lambda_F$ is the interaction range, e.g., the Thomas-Fermi interaction in Eqs.~(\ref{VTFr}), (\ref{VTFq}).
The asymptotics of $\chi_{s's}^{(1)}\left(Q_{s's}^+, r\right)$ in Eq.~(\ref{chi2kFr}), at $r \gg R_0$ can be derived for a general finite-range interaction.
Indeed, the integral over $y$ in Eq.~(\ref{chi2kFr}) is convergent on the scale of $|y| \lesssim R_0/r \ll 1$ which allows us to extend the integration over $y$ all the way to $-\infty$ and to use $|y| \ll 1$ asymptotics of $\mathcal{F}^+(y)$ in Eq.~(\ref{Fplus}),
\begin{eqnarray}
&& \mathcal{F}^+(y) = -\frac{1}{y} - \ln|y| + \ln 2 - \frac{1}{2} + \mathcal{O}(y) . \label{Fsmall}
\end{eqnarray}
As $V(x)$ is an even function, the leading $-1/y$ term in Eq.~(\ref{Fsmall}) does not contribute to the integral in Eq.~(\ref{chi2kFr}).
Therefore, the leading approximation of the integral in Eq.~(\ref{chi2kFr}) is the following
\begin{eqnarray}
&& \hspace{-13pt} \int\limits_{-1}^{+\infty} dy \, V_{R_0}(r y) \mathcal{F}^+(y)  \approx -\int\limits_{-\infty}^{+\infty} dy \, V_{R_0}(ry) \left[\ln\left|\frac{y}{2}\right|+ \frac{1}{2}\right]  \nonumber \\
&& = \frac{2}{r} \left[V_1 + V_2 \left(\ln\left(\frac{2 r}{R_0}\right)- \frac{1}{2} \right)\right] , \label{zint}
\end{eqnarray}
where we introduced the following constants,
\begin{eqnarray}
&& \hspace{-20pt} V_1 = -\int\limits_{0}^\infty dx \, V_{R_0}(x) \ln \left|\frac{x}{R_0}\right| \nonumber \\
&& \hspace{-7pt} = \int\limits_0^\infty \frac{dq}{2 \pi} \, V_{R_0}(q) \left[\ln \left(2 q R_0\right) + \gamma\right] , \label{V1} \\
&& \hspace{-20pt} V_2 = \int\limits_{0}^\infty dx \, V_{R_0}(x) = \int\limits_0^\infty \frac{dq}{2\pi} \, V_{R_0}(q). \label{V2}
\end{eqnarray}
Here, $V_{R_0}(q)$ is the 2D Fourier transform of $V_{R_0}(x)$,
$\gamma \approx 0.577$ is the Euler-Mascheroni constant, and we have used the fact that $V_{R_0}(x)$ must be an even function.
Notice that both $V_1$ and $V_2$ are positive constants for an arbitrary finite-range repulsive interaction.
Substituting Eq.~(\ref{zint}) into Eq.~(\ref{chi2kFr}), we find the $Q_{s's}^+ \approx 2 k_F$ harmonic of the static susceptibilities for an arbitrary finite-range interaction
\begin{eqnarray}
&& \hspace{-15pt} \chi_{s's}^{(1)}\left(Q_{s's}^+, r\right) \nonumber \\
&& \approx - \frac{N_F V_2}{2 \pi^2 v_F} \frac{\sin\left(Q_{s's}^+ r\right)}{r^2} \left[\frac{1}{2} - \ln \left|\frac{2 r}{R_0}\right| - \frac{V_1}{V_2}\right] , \label{chi2kFR0r}
\end{eqnarray}
where $N_F$ is the density of states [see Eq.~(\ref{NF})], and $R_0 \gg \lambda_F$ is the interaction range.
The asymptotics given by Eq.~(\ref{chi2kFR0r}) is valid at $r \gg R_0$.
The Fourier transform of this expression (see Appendix~\ref{app:Fourier}) results in the following two-sided square-root non-analyticity
\begin{eqnarray}
&& \hspace{-10pt} \chi_{s's}^{(1)}\left(Q_{s's}^+, q\right) \approx - \frac{\sqrt{2} N_F V_2}{2 \pi v_F} \sqrt{\left|\frac{\delta q}{Q_{s's}^+}\right|} \left\{ \vphantom{\frac{1}{2}} \pi \, \vartheta\left(-\delta q\right)
\right. \nonumber \\
&& \hspace{10pt} \left. + 2 \, \vartheta  \left(\delta q\right) \left[\frac{V_1}{V_2} -  \ln\left|2 \, \delta q \, R_0\right| + \frac{3}{2} - \gamma \right] \right\} , \label{chi2kFR0q}
\end{eqnarray}
where $\delta q = q - Q_{s's}^+ \ll Q_{s's}^+$. 
The asymptotics given by Eq.~(\ref{chi2kFR0q}) is valid for any finite-range interaction at $\delta q \ll 1/R_0 \ll k_F$.
As we clearly see here, the non-analyticity survives for an arbitrary finite-range interaction, it is two-sided, negative and enhanced by the logarithm at $q > Q_{s's}^+ \approx 2 k_F$.
Such a non-analyticity will always result in a peak structure near the $2 k_F$ Kohn anomaly.

As a concrete example of a finite-range interaction, we now consider the Thomas-Fermi interaction given by Eq.~(\ref{VTFq}).
The constants $V_1$ and $V_2$ for the Thomas-Fermi interaction are the following (see Appendix~\ref{app:TF}):
\begin{eqnarray}
&& V_1^{TF} \approx \frac{1}{m a_B} \ln\left(\frac{k_\Lambda}{\kappa}\right) \ln\left(2 e^\gamma R_0 \sqrt{\kappa k_\Lambda}\right) , \label{V1TF} \\
&& V_2^{TF} \approx \frac{1}{m a_B} \ln \left(\frac{k_\Lambda}{\kappa}\right) , \label{V2TF}
\end{eqnarray}
where $a_B$ and $\kappa$ are defined in Eq.~(\ref{kappa}).
To evaluate these interaction constants, the integrals over $q$ in Eqs.~(\ref{V1}), (\ref{V2}) were taken over the interval $q \in (0, k_\Lambda)$, where $k_\Lambda \sim k_F \gg \kappa$ is the ultraviolet cut-off of the semiclassical approximation.
Substituting Eqs.~(\ref{V1TF}), (\ref{V2TF}) into Eq.~(\ref{chi2kFR0q}), we find
\begin{eqnarray}
&& \hspace{-23pt} \chi_{s's}^{(1)}\left(Q_{s's}^+, q\right) \approx - \frac{r_s N_F}{\pi} \sqrt{\left|\frac{\delta q}{Q_{s's}^+}\right|} \ln \left(\frac{k_\Lambda}{\kappa}\right) \nonumber\\
&& \hspace{34pt} \times \left[\vartheta\left(\delta q\right) \ln\left|\frac{e^{\frac{3}{2}} \sqrt{\kappa k_\Lambda}}{\delta q}\right| + \frac{\pi}{2} \vartheta\left(-\delta q\right) \right] ,  \label{chiTF}
\end{eqnarray}
where $\delta q = q - Q_{s's}^+ \ll \kappa \ll k_F$.
The non-analyticity in Eq.~(\ref{chiTF}) has a resonant peak structure near $q = Q_{s's}^+ \approx 2 k_F$
that was first seen numerically in Ref.~\cite{ashcroft} for the Coulomb interaction dynamically screened by the particle-hole bubble.

The semiclassical method of the dimensional reduction \cite{mis2023} allowed us to analytically find the strong resonant contribution to the susceptibilities that originates from the scattering processes near the FS.
Here we stress again that Eq.~(\ref{chiTF}) represents the asymptotics at $|q - Q_{s's}^+| \ll \kappa \ll k_F \sim k_\Lambda$.
This asymptotics crosses over to Eq.~(\ref{chi2kFq}) that is valid at $\kappa \ll |q - Q_{s's}^+| \ll k_F$.
Notice that the prefactor in Eq.~(\ref{chiTF}) contains $\ln(k_\Lambda/\kappa)$, which suppresses the peak at $\kappa \approx k_\Lambda \sim k_F$.
This corresponds to the limit of the contact or zero-range interaction, where no resonant peak structure of the susceptibilities at the $2 k_F$ Kohn anomaly is expected, see e.g. Refs.~\cite{lohneysen,vojta,belitzkirk1997,chubmaslov2003,chubglazman2005,chesizak,zak2012,maslov,zak2010,mis2019,mis2021,mis2022}, which is therefore consistent with our results.
It is worth noting here that the perturbation theory breaks down at $r_s \gtrsim 1$.
The regime $r_s \gtrsim 1$ requires a non-perturbative treatment which goes beyond the scope of this work.
Nevertheless, we believe that our results remain qualitatively correct even at $r_s \sim 1$.

\section{Full Susceptibilities}\label{sec:suscept}

In this section we sum up the Dyson series for the full static susceptibility and approximate the irreducible susceptibility by the sum of zero-order and first-order irreducible susceptibilities shown in Figs.~\ref{fig:free}, \ref{fig:firstorder}.
Inclusion of the first-order corrections to the irreducible susceptibilities is a step beyond the random phase approximation.
Here we show that the charge susceptibility is suppressed by the electrostatic screening compared to the spin susceptibility.
In other words, a repulsive electron-electron interaction favors the formation of $2 k_F$ SDW orders rather than the $2 k_F$ charge DW (CDW) order in the interacting 2DEG.
It is worth noting that such summation was not performed in Ref.~\cite{ashcroft} which resulted in largely overestimated $2 k_F$ peak of the charge susceptibility.

The Dyson equation for the 2D static susceptibilities can be represented as follows
\begin{eqnarray}
&& \hspace{-15pt} \chi_{ij} (q) = \chi_{ij}^{\rm irr}(q) - \chi^{\rm irr}_{i0}(q) \tilde V(q) \chi_{0j}^{\rm irr}(q) , \label{chiij} \\
&& \hspace{-15pt} \tilde V(q) = \frac{V_C(q)}{1 + \chi_{00}^{\rm irr}(q) V_C(q)} , \label{Vdressed}
\end{eqnarray}
where $\chi^{\rm irr}_{ij}(q)$ is the irreducible static susceptibility, indices $i, j \in \{0,1,2,3\}$ label the spin matrices $\eta_i$ and $\eta_j$ introduced in Eqs.~(\ref{eta1})--(\ref{eta3}) (with $\eta_0 = I$, the identity matrix).
Here, $\tilde V(q)$ is the dressed electron-electron interaction and $V_C(q)$ is the Coulomb interaction [see Eq.~(\ref{Coulomb})]. The polarization operator is equal to $-\chi_{00}^{\rm irr}(q)$.
Here we approximate the irreducible susceptibilities by the free-fermion susceptibility, plus the first-order diagrams (see Figs.~\ref{fig:free} and~\ref{fig:firstorder}),
\begin{eqnarray}
&& \chi_{ij}^{\rm irr}(q) \approx \chi_{ij}^{(0)}(q) + \chi_{ij}^{(1)}(q) = \sum\limits_{s, s'} \eta_i^{s s'} \eta_j^{s's} \chi_{s's}^{\rm irr} (q) , \label{chiirr}\\
&& \chi_{s' s}^{\rm irr}(q) = \chi_{s's}^{(0)}(q) + \chi_{s's}^{(1)}(q) ,  \label{chiirrss}
\end{eqnarray}
where $\chi_{s's}^{(0)}(q)$ is given by Eq.~(\ref{chiq0}), and $\chi_{s's}^{(1)}(q)$ is considered in Sec.~\ref{sec:firstorder} for different interactions. 
In particular, $\chi_{s's}^{(1)}(q)$ is given by Eq.~(\ref{chiTF}) at $|q - Q_{s's}^+| \ll \kappa \ll k_F$ and by Eq.~(\ref{chi2kFq}) at $\kappa \ll |q - Q_{s's}^+| \ll k_F$.
These two limits describe the largest anomaly in $\chi_{s's}^{(1)}(q)$ that is located in the vicinity of $q = Q_{s's}^+ = k_{s'} + k_s \approx 2 k_F$.

\begin{figure}
	\includegraphics[width=0.98\columnwidth]{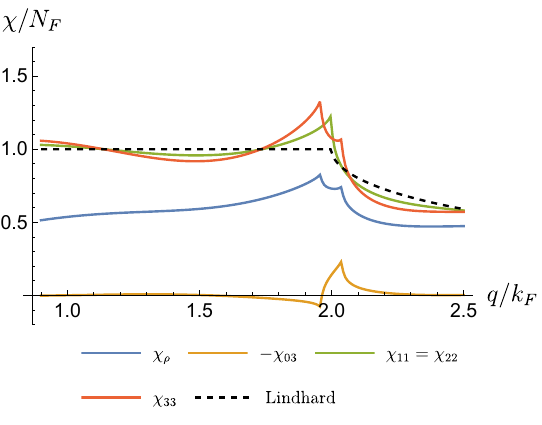}
	\caption{Components of the full susceptibilities near $q = 2 k_F$ [see Eqs.~(\ref{chirho})--(\ref{chi11})] at $\kappa = 0.2 k_F$, $r_s = 1.0$ and $\delta = 0.02 k_F$.
	Irreducible susceptibilities are calculated numerically as a sum of diagrams in Figs.~\ref{fig:free}, \ref{fig:firstorder} with the Thomas-Fermi interaction [see Eq.~(\ref{VTFq})].
	Susceptibilities $\chi_\rho(q)$, $\chi_{33}(q)$, $-\chi_{03}(q)$ have two peaks at $q = 2 k_-$, $q = 2 k_+$; $\chi_{11}(q)$ has single peak at $q = k_+ + k_- = 2 k_F$, see Eq.~(\ref{ks}).
	The spin susceptibility $\chi_{33}(q)$ demonstrates the largest peak at $q = 2 k_-$.
	The charge susceptibility $\chi_\rho (q)$ is suppressed by the electrostatic screening.
	For reference, the non-interacting Lindhard function $\chi^{(0)}_{11}(q)$, see Eq.~(\ref{chiq0}), is shown by black dashed curve.}
	\label{fig:chiscreen}
\end{figure}

\begin{figure}
	\includegraphics[width=0.98\columnwidth]{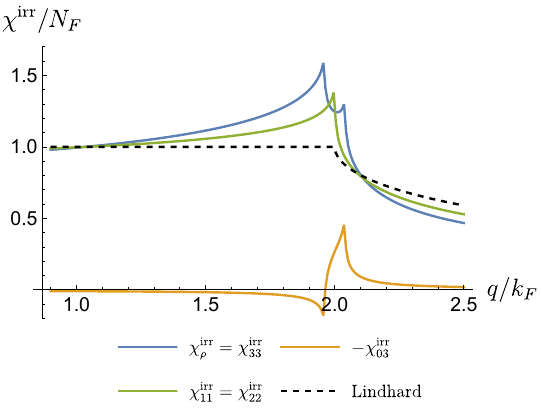}
	\caption{Components of the irreducible susceptibilities given by the sum of diagrams in Figs.~\ref{fig:free}, \ref{fig:firstorder} near $q = 2 k_F$ for the unscreened Coulomb interaction [Eq.~(\ref{Coulomb})] with $r_s = 1.0$ and $\delta = 0.02 k_F$. Note from Eq.~\eqref{chiirrss} that $\chi_\rho^{\rm irr}=\chi_{33}^{\rm irr}$, which is shown by the blue curve.
	The non-interacting Lindhard function $\chi^{(0)}_{11}(q)$, see Eq.~(\ref{chiq0}), is shown by black dashed curve.}
	\label{fig:chi}
\end{figure}

First, we notice that $\chi_{0i}^{\rm irr}(q) = \chi_{3 i}^{\rm irr}(q) = 0$ for $i \in \{1, 2\}$ which follows from Eq.~(\ref{chiirr}) and $\eta_1^{ss'} = \eta_2^{ss'} = 0$ at $s' = s$, where $\eta_i^{ss'} = \langle \psi_s| \eta_i | \psi_{s'} \rangle$.
Similarly, one can directly check that $\chi_{12}^{\rm irr}(q) = 0$.
Therefore, all non-zero components of the full static susceptibility are the following:
\begin{eqnarray}
&& \hspace{-25pt} \chi_\rho (q) \equiv \chi_{00}(q) = \frac{\chi_{++}^{\rm irr}(q) + \chi_{--}^{\rm irr}(q)}{1 + V_C(q) \left[\chi_{++}^{\rm irr}(q) + \chi_{--}^{\rm irr}(q) \right]} , \label{chirho} \\
&& \hspace{-25pt} \chi_{30}(q) = \chi_{03}(q) = \frac{\chi_{++}^{\rm irr}(q) - \chi_{--}^{\rm irr}(q) }{1 + V_C(q) \left[\chi_{++}^{\rm irr}(q) + \chi_{--}^{\rm irr}(q) \right]} , \label{30} \\
&& \hspace{-25pt} \chi_{33}(q) = \chi_\rho (q) + 4 \tilde{V} (q) \chi_{++}^{\rm irr}(q) \chi_{--}^{\rm irr}(q) , \label{33} \\
&& \hspace{-25pt} \chi_{11} (q) = \chi_{22} (q) = 2 \chi_{+-}^{\rm irr} (q) , \label{chi11} 
\end{eqnarray}
where $\chi_\rho(q)$ is the charge susceptibility, $\tilde{V}(q)$ is the dressed interaction [see Eq.~(\ref{Vdressed})], $\chi_{ss'}^{\rm irr}(q)$ are defined in Eq.~(\ref{chiirrss}), where $s, s' \in \{\pm 1\}$.
As we see from equations above, $\chi_{33}(q) > \chi_\rho(q)$ at any $q$, i.e.,
the spin density correlations are always dominant in an interacting 2DEG.
We also notice that $\chi_{11}(q) = \chi_{22}(q)$ remain unscreened.
Different non-zero components of susceptibilities are plotted in Fig.~\ref{fig:chiscreen}, where the screening effects are included via Eqs.~(\ref{chirho})--(\ref{chi11}) and via the Thomas-Fermi screening of the interaction in diagrams in Fig.~\ref{fig:firstorder}.
It is clear from Fig.~\ref{fig:chiscreen} that the largest susceptibility component is the spin susceptibility $\chi_{33}(q)$ at $q = 2 k_-$ which confirms our expectations that electrostatic screening suppresses the charge susceptibility compared to the spin susceptibility.
In order to demonstrate the effect of electrostatic screening, we also plot the corresponding components of the unscreened irreducible susceptibilities in Fig.~\ref{fig:chi}. These components are represented by the sum of diagrams in Figs.~\ref{fig:free} and \ref{fig:firstorder} calculated for the unscreened Coulomb interaction from Eqs.~\eqref{static2D},~\eqref{chisumr}-\eqref{a} and then Fourier transformed numerically.

We should note that the susceptibility components given in Eqs.~(\ref{chirho})--(\ref{chi11}) are written in the rotated basis [see Eqs.~\eqref{eta1}-\eqref{eta3}].
In order to get the physical responses to different perturbations, one should rotate back.
For example, the true spin susceptibility, i.e. the response in the spin-$z$ channel to a magnetic field perpendicular to the 2DEG plane is given by
\begin{equation}
\chi_{zz}(q) = \sin^2\theta \, \chi_{11}(q) +\cos^2\theta \, \chi_{33}(q) , \label{zz}
\end{equation}
while the mixed spin-charge susceptibility, i.e. the spin-$z$ response to an electric potential perturbation is
\begin{equation}
\chi_{0z} (q) = -\cos\theta \, \chi_{03}(q) .\label{eq:chi0z}
\end{equation}
Here, we used that $\chi_{13}(q) = \chi_{31}(q) = \chi_{01}(q) = 0$.
For example, the red (green) curve in Fig.~\ref{fig:chiscreen} can be understood as $\chi_{zz}(q)$ at $\theta = 0$ ($\theta = \pi/2$), while the orange curve coincides with $\chi_{0z}(q)$ at $\theta = 0$. 
In a bilayer system, $\chi_{zz}(q)$ represents the out-of-plane electric dipole response to a local potential difference across the two layers.  
In this case, $\theta = 0$ ($\theta = \pi/2$) corresponds to $\alpha \gg t_\perp$ ($\alpha \ll t_\perp$), where $\alpha$ is the layer bias, $t_\perp$ is the layer hybridization, see Eqs.~(\ref{Bbi}) and (\ref{theta}).
Figures~\ref{fig:chiscreen} and \ref{fig:chi} and Eq.~(\ref{zz}) demonstrate that tuning $\theta$ shifts the Kohn anomaly from the double-peak at $q = 2 k_F \pm 2 \delta$ to a single peak at $q = 2 k_F$.

The difference between Eq.~\eqref{chirho} and Eq.~\eqref{33} clearly shows that $\chi_{33}(q) > \chi_\rho(q)$, i.e., the 2DEG is more susceptible to the SDW order than to the CDW order. 
In a bilayer system, this means that density waves consisting of interlayer dipoles are expected to be far more prevalent than a standard CDW.
This behaviour is expected because of the large electrostatic energy needed to stabilize the CDW order. 
Meanwhile the SDW order does not create any local charge imbalance.

\section{Susceptibilities of a 3DEG}
\label{sec:3DEG}

In this section, we wish to make a comparison between the two- and the three-dimensional electron gas (3DEG), also in order to underline the special features emerging in lower dimensions.
In particular, the 3D free-fermion susceptibility is given by the well-known expression (see Refs.~\cite{kohn1959,kohn1962,afanasev} and Appendix~\ref{app:free}).
\begin{eqnarray}
&& \hspace{-15pt}\chi_{s's}^{(0)} (q) = \frac{N_F Q_{s's}^+}{8 k_F q^2} \left[q^2 + Q_{s's}^{-2} \right. \nonumber \\
&& \hspace{17pt} \left. + \frac{\left(q^2 - Q_{s's}^{-2}\right) \left(Q_{s's}^{+2} - q^2\right)}{2 q Q_{s's}^+} \ln\left|\frac{q + Q_{s's}^+}{q - Q_{s's}^-}\right|\right] , \label{chi03D}
\end{eqnarray}
where the components $\chi_{s's}^{(0)}(q)$ are defined in the same way as in Eq.~(\ref{chiq0}), the $Q_{s's}^\pm$ components are defined in Eq.~(\ref{Qpm}), $N_F = m \mathfrak{g} k_F/\pi^2$ is the total 3D density of states at the Fermi energy, and the 3D Hamiltonian is assumed to have the form of Eq.~(\ref{eq:interhop}) but with a 3D momentum.
The free-fermion 3D susceptibility $\chi_{s's}^{(0)} (q)$ is analytic everywhere except at $q = Q_{s's}^+ \approx 2 k_F$.
The main difference with respect to a 2DEG here is that $\chi_{s's}^{(0)}(q)$ does not have a plateau at $q < Q_{s's}^+$; its maximal value corresponds to $q = 0$ and it drops down by a factor of 2 (at $Q_{s's}^- = 0$) at $q = Q_{s's}^+$. 
Therefore, small interaction corrections near the Kohn anomaly at $q = Q_{s's}^+$ are not as important as in a 2DEG.
Instead, the maximum of the susceptibility near $q = 0$ may lead to a uniform ferromagnetic instability; this effect has been studied extensively, see e.g., Refs.~\cite{vojta,lohneysen}.

We note that the first-order non-analytic correction to the susceptibility of a 3DEG can be calculated using the one-dimensional susceptibility given by Eq.~(\ref{chisumr}).
Applying the dimensional reduction technique \cite{mis2023} to a 3DEG, one can easily find the long-distance asymptotics of static 3D susceptibilities that is analogous to the 2D case, see Eq.~(\ref{static2D}),
\begin{eqnarray}
&& \chi_{s's}^{(1)}(r) = \mathfrak{g} \sum\limits_{\nu, \nu'} \frac{e^{i \frac{\pi}{2}(\nu - \nu')}}{\lambda_s \lambda_{s'}} \frac{e^{i r Q_{\gamma'\gamma}}}{r^2} \tilde \chi^{(1)}(Q_{\gamma' \gamma}, r) , \label{chi3Dr}
\end{eqnarray}
where $\tilde \chi^{(1)}(Q_{\gamma' \gamma}, r)$ is given by Eq.~(\ref{chisumr}).
The 3D Fourier transform of Eq.~(\ref{chi3Dr}) at $q \approx Q_{s's}^+$ yields the non-analyticity $\propto \delta q [\ln|\delta q|]^2$, with $\delta q = q - Q_{s's}^+$, for an arbitrary finite-range interaction, which is the same non-analyticity that is already present in the free-fermion susceptibility given by Eq.~(\ref{chi03D}) with an additional power of the logarithm.
This is why interaction effects on the susceptibility of a 3DEG near the $2 k_F$ Kohn anomaly are rather marginal which has also been pointed out in Ref.~\cite{ashcroft}.

Even though the Kohn anomaly of an isotropic 3DEG seems to be unimportant for magnetic instabilities, things might change in  3D layered van der Waals materials where the interlayer hopping might be orders of magnitude smaller than the intralayer one.
In this case we expect a crossover to the 2D case where the free-fermion susceptibility flattens out at $q < 2 k_F$, thus making the interaction effects more prominent near $q = 2 k_F$.

\section{Discussion of $2 k_F$ instabilities}
\label{sec:insta}

Here we discuss and explore possible physical consequences of sharp $2 k_F$ resonances of the 2D static susceptibilities.
In particular, we discuss possible intrinsic SDW instabilities, magnetic ordering of ensembles of magnetic impurities and nuclear spins via the RKKY interaction~\cite{ruderman,kasuya,yosida}, and instabilities of the crystal order via the electron-phonon coupling.

Even with finite 2D static spin susceptibilities, there are a number of phase transitions that can be induced by the resonant $2 k_F$ Kohn anomaly found in this work. 
First example is related to the RKKY interaction~\cite{ruderman,kasuya,yosida} describing effective spin-spin coupling between localized magnetic moments embedded in an electron gas whose electrons interact with the local moments via finite exchange interaction $J$. 
To the lowest order in $J$, the resulting RKKY interaction is determined by the spin susceptibility of the electron gas.
In particular, a magnetic {\it helical ordering} of the nuclear spins or magnetic impurities is possible due to the spin-density fluctuations in the 2DEG that are enhanced by electron-electron interactions.
Indeed, if the RKKY coupling is sufficiently strong (larger than some finite critical value), local magnetic moments embedded in the interacting 2DEG will order into a $2 k_F$ helical ground state.
Such magnetic ordering creates an effective magnetic field oscillating in space with the $2 k_F$ wave vector which results in a helical gap opening in the electron spectrum, see e.g. Refs.~\cite{braunecker,japaridze} for the concrete physical mechanism and the caveats related to the generalized Mermin-Wagner theorem~\cite{loss_leggett}.
Moreover, proximity coupling to a superconductor may further result in the induced topological superconductivity, see e.g. Refs.~\cite{klinovajastano,perge2013,brauneckersimon2013}.

\begin{figure}
	\includegraphics[width=0.88\columnwidth]{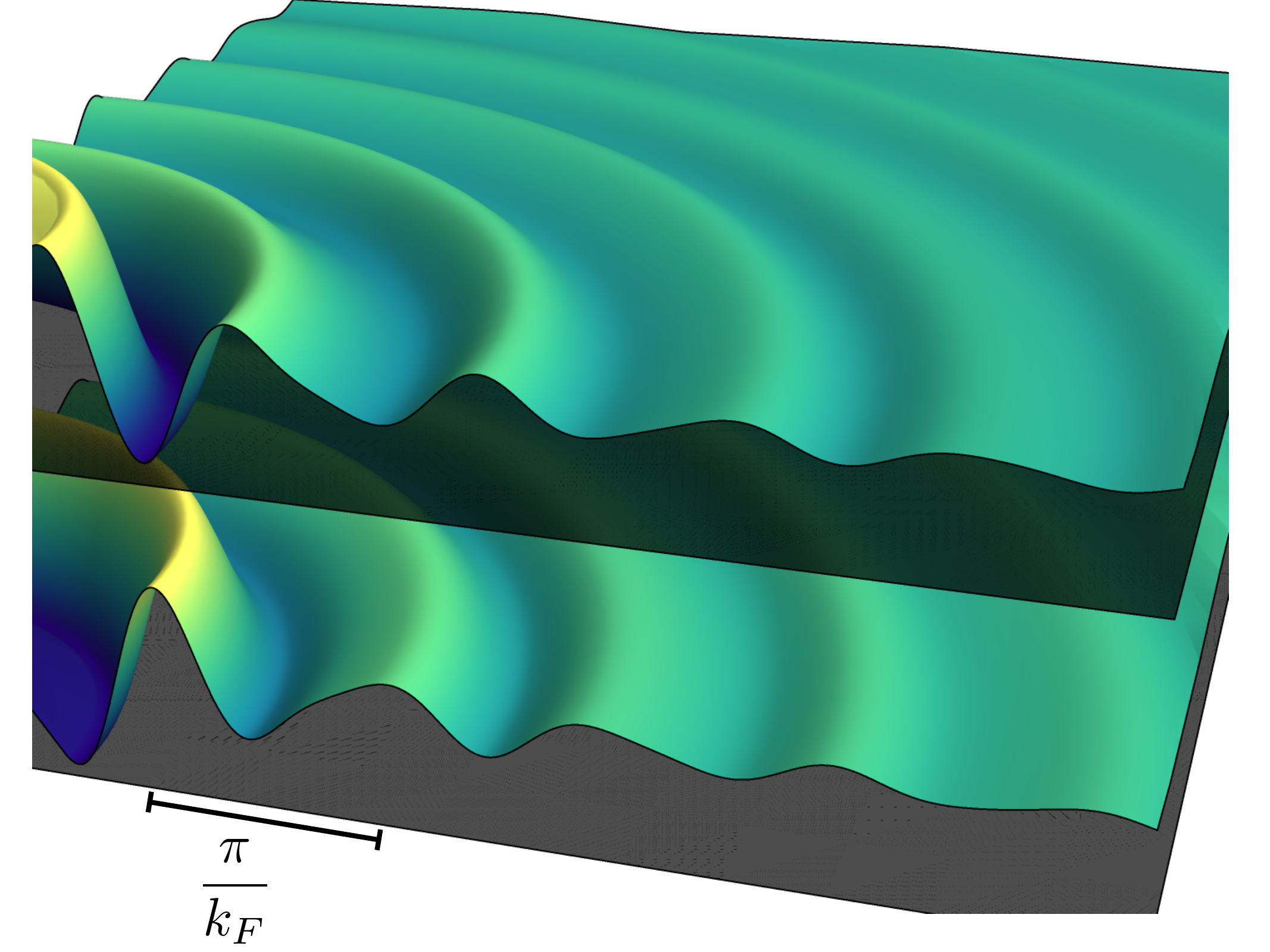}
	\caption{Density response in each layer of a bilayer 2DEG to a local potential difference $\phi\delta(\b{r})$. The two layers have decaying Friedel oscillations of wavelength $\pi/k_F$ exactly out of phase with each other.}
	\label{fig:cdw}
\end{figure}

In this paper, we also emphasize that the same resonant peaks at the $2 k_F$ Kohn anomaly appear in the layer pseudospin susceptibilities in bilayer materials such as bilayer graphene~\cite{mccann,rozhkov,szabo} and bilayer TMDs~\cite{yankowitz2015,xu2021,chen2016,kidd2002,leroux2015,zhang2022,zhao2022}. In this case, the pseudospin susceptibilities derived here provide the linear charge response in each layer ($\delta n_l$), to perturbations in the electric potential $\phi_l$ via
\be
\begin{pmatrix}
\delta n_1(\b{r})\\
\delta n_2(\b{r})
\end{pmatrix} = \int d^2r'\, \Gamma\chi(\b{r}-\b{r}')\Gamma
\begin{pmatrix}
\phi_1(\b{r}')\\
\phi_2(\b{r}')
\end{pmatrix},
\ee  
where the susceptibility $\chi$ is written in the un-rotated basis, and $\Gamma$ transforms this to the layer basis,
\be
\chi=
\begin{pmatrix}
\chi_\rho & \chi_{0z} \\
\chi_{0z} & \chi_{zz}
\end{pmatrix},\;\;\Gamma=\frac{1}{\sqrt{2}}
\begin{pmatrix}
1 & 1 \\
1 & -1
\end{pmatrix}.
\ee
Consider a local potential difference $\phi_1(\b{r})\equiv\phi\delta(\b{r})=-\phi_2(\b{r})$. Noting from Fig.~\ref{fig:chiscreen} and Eq.~\eqref{eq:chi0z} that $\chi_{0z}$ is much smaller than the other components, the charge response is simply given by the $zz$-component of the pseudospin susceptibility, i.e.
\be
\delta n_1(\b{r})=-\delta n_2(\b{r})=\phi\chi_{zz}(\b{r}).
\ee
These density perturbations are shown in Fig.~\ref{fig:cdw}. In general, local impurities will induce oscillations at wavenumber $2k_F$ with a $\pi$ phase difference between the layers, forming a dipole density wave.

As in the case with magnetic impurities and nuclear spins, a true dipole-density-wave instability may be instigated by coupling to phonons: at large enough electron-phonon coupling or soft enough phonon modes, the Kohn anomaly can result in phonon condensation, causing an effective $2 k_F$-periodic potential with a relative $\pi$ phase-shift between the layers.
This potential opens a partial gap in the electron spectrum. 
In principle, the $2 k_F$ dipole density wave together with the interlayer bias and proximity to an $s$-wave superconductor may be also considered as an alternative way to engineer topological superconductors.

\section{Conclusions}
\label{sec:conc}

In this paper we performed a calculation of the spin susceptibility of an isotropic 2DEG with long-range and finite-range interactions at first-order in perturbation theory.
As the 2D free-fermion susceptibilities have a plateau at $q < 2 k_F$, the interaction effects become particularly important.
Here we find that already the first-order interaction contribution to the spin susceptibility results in a resonant peak structure near the $2 k_F$ Kohn anomaly for a Coulomb interaction [see Eq.~(\ref{chi2kFq})], as well as for an arbitrary finite-range repulsive interaction [see Eq.~(\ref{chi2kFR0q})]. This is of fundamental importance for the spin physics and magnetic phases of low-dimensional systems.
Our analytic calculation reveals a structure of resonant Kohn anomaly in the 2D static susceptibilities of both charge and spin. 
While for the charge susceptibility this has been previously spotted numerically in Ref.~\cite{ashcroft}, the spin susceptibility has not been analyzed before for finite-range interactions.
We also show that the charge susceptibility is suppressed compared to the spin susceptibility due to the electrostatic screening.

Our analysis is further applicable to bilayer materials, where the layer index plays the role of a pseudospin. 
In particular, we find that the dominant linear response is the out-of-plane dipole susceptibility, $\chi_{zz}$.
When the electron-phonon coupling is introduced, the resonant Kohn anomaly in $\chi_{zz}$ can result in an instability towards a $2k_F$ density wave of interlayer dipoles. 
Such a dipole-density wave is equivalent to two $2 k_F$ CDWs in each layer with the relative phase shift of $\pi$. 
The dipole-density wave can be observed experimentally by local probes such as STM and nitrogen-vacancy centers.

We also argue that peaked spin susceptibilities at $2 k_F$ may instigate magnetic helical order of localized spins via the RKKY interaction.

Finally, we note that the logarithmic corrections to the 2D susceptibilities which we derived in this manuscript signal a strong renormalization of the bare square-root non-analyticity near the $2 k_F$ Kohn anomaly.
Summation over the leading logarithmic corrections at all orders of perturbation theory may further clarify whether intrinsic SDW instabilities are possible in an interacting 2DEG or not.
The construction of the renormalization group procedure for an interacting 2DEG with a finite-range repulsive interaction is a subject of our future study.

\begin{acknowledgments}
This project has received funding from the European Union's Horizon 2020 research and innovation programme under Grant Agreement No 862046 and under  Grant Agreement No 757725 (the ERC Starting Grant). This work was supported by the Georg H. Endress Foundation and the Swiss National Science Foundation.
\end{acknowledgments}

\appendix

\section{Free-fermion 2D and 3D susceptibilities}
\label{app:free}

Here we briefly review well-known results for the free-fermion 2D and 3D susceptibilities \cite{kohn1959,kohn1962,afanasev} generalized for the split FS.
The free-fermion static susceptibilities of a $D$-dimensional electron gas are given by the diagram in Fig.~\ref{fig:free}
\begin{eqnarray}
&& \hspace{-40pt} \chi^{(0)}_{s's} (q) = - \mathfrak{g} \int\frac{d \bm p}{(2 \pi)^D} \nonumber \\
&& \hspace{10pt} \times  \int\limits_{-\infty}^\infty \frac{d \omega}{2 \pi} G_s (\bm p, i \omega) G_{s'} (\bm p + \bm q, i \omega) , \label{A:chi0D}
\end{eqnarray}
where $\bm p$ and $\bm q$ are the $D$-dimensional momenta, the temperature is set to zero, and the free-fermion Green's function is represented by the quasiparticle pole,
\begin{eqnarray}
&& G_s (\bm p, i \omega) = \frac{1}{i \omega - \varepsilon_s (p)} . \label{A:freegreen}
\end{eqnarray}
The electron dispersion $\varepsilon_s(p)$ is given by Eq.~(\ref{spec}).
The integral over frequency yields the standard Lindhard function
\begin{eqnarray}
&& \hspace{-18pt} \chi_{s's}^{(0)}(q) = - \mathfrak{g} \! \int \!\! \frac{d \bm p}{(2 \pi)^D} \frac{\vartheta\left(k_{s'} - |\bm p + \bm q|\right) - \vartheta\left(k_{s} - p\right)}{\varepsilon_{s'} (\bm p + \bm q) - \varepsilon_s (p)} , \label{A:lindhard}
\end{eqnarray}
where $\vartheta(x)$ is the Heaviside step function, $\vartheta(k_s - p)$ represents the Fermi-Dirac distribution function at zero temperature, while $k_s$ is the Fermi momentum of the $s^{\rm th}$ FS [see Eq.~(\ref{ks})]. 
It is more convenient to introduce an auxiliary function $f_{s's}(q)$
\begin{eqnarray}
&& \chi_{s's}^{(0)}(q) = f_{s's}(q) + f_{ss'}(q), \label{A:chif} \\
&& f_{s's}(q) = - \mathfrak{g} \int \frac{d \bm p}{(2 \pi)^D} \frac{\vartheta(k_{s'} - p)}{\varepsilon_{s'}(p) - \varepsilon_s(\bm p + \bm q)} . \label{A:f}
\end{eqnarray}
Here all integrations will be performed for the parabolic electron dispersion [see Eq.~(\ref{spec}))] not for the linearized one.

First, we consider the 2D case. For this, we require the following angular integral
\begin{eqnarray}
&& \int\limits_0^{2 \pi} \frac{d \phi}{2 \pi} \frac{1}{\cos \phi + z + i 0 \sigma} = \frac{\sgn(z)}{\sqrt{\left(z + i 0 \sigma\right)^2 - 1}} , \label{A:intphi2D}
\end{eqnarray}
where $\sigma = \pm 1$, $z$ is an arbitrary real number, $\sgn(z)$ returns the sign of $z$, the square root represents the principal branch, i.e., its real part is non-negative, and the infinitesimal term $i 0 \sigma$ is added to avoid problems at $-1 < z < 1$.
The easiest way to derive Eq.~(\ref{A:intphi2D}) is to introduce a complex variable $e^{i \phi}$ such that the initial integral is reduced to a complex integral from a meromorphic function over the unit circle.
Using Eq.~(\ref{A:intphi2D}), we perform the integration over the angle $\phi$ between $\bm p$ and $\bm q$ in Eq.~(\ref{A:f}),
\begin{eqnarray}
&& f_{s's}(q) = \frac{N_F}{2 q}  \int\limits_0^{k_{s'}} \frac{\sgn\left[\mu_{s's}(q)\right] \, dp}{\sqrt{\left(\frac{\mu_{s's}(q)}{p} + i 0 \sigma\right)^2 - 1}} , \label{A:fp} \\
&& \mu_{s's}(q) = \frac{q}{2} + \frac{k_{s'}^2 - k_s^2}{2 q} , \label{A:mu}
\end{eqnarray}
where $N_F$ is the 2D density of states [see Eq.~(\ref{NF})], $\sigma = \pm 1$ is chosen arbitrarily; the final result for $\chi_{s's}(q)$ is independent of the choice of $\sigma$.
The remaining integral over $p$ in Eq.~(\ref{A:fp}) is elementary
\begin{eqnarray}
&& \hspace{-20pt} f_{s's}(q) = \frac{N_F}{2 q} \left[ \vphantom{\sqrt{\left(\mu_{s's}(q) + i 0 \sigma\right)^2 - k_{s'}^2}} \mu_{s's}(q) \right. \nonumber \\
&& \hspace{10pt} \left. - \sgn\left(\mu_{s's}(q)\right) \sqrt{\left(\mu_{s's}(q) + i 0 \sigma\right)^2 - k_{s'}^2} \hspace{2pt}\right] , \label{A:f2D}
\end{eqnarray}
where the square root represents the principal branch.
Substituting Eq.~(\ref{A:f2D}) into Eq.~(\ref{A:chif}) and performing elementary algebraic transformations, we find the free-fermion 2D susceptibility given by Eq.~(\ref{chiq0}).

The angular integration in 3D case is even simpler,
\begin{eqnarray}
&& \int\limits_0^{2\pi} d\phi \int\limits_0^\pi \frac{\sin \beta \, d\beta}{\cos \beta + z} = 2 \pi \ln \left|\frac{z + 1}{z - 1}\right| , \label{A:ang3D}
\end{eqnarray}
where $z$ is an arbitrary real number, $\phi$ and $\beta$ are 3D azimuthal and polar angles, respectively.
Choosing the polar direction along $\bm q$ and introducing the angles $\phi$ and $\beta$ defining the position of $\bm p$ in Eq.~(\ref{A:f}), we perform the angular integration via Eq.~(\ref{A:ang3D}),
\begin{eqnarray}
&& f_{s's}(q) = \frac{N_F}{4 q k_F} \int\limits_0^{k_{s'}} \ln\left|\frac{p + \mu_{s's}(q)}{p - \mu_{s's}(q)}\right| p \, dp , \label{A:fp3D}
\end{eqnarray}
where we introduced the 3D density of states $N_F = m \mathfrak{g} k_F/\pi^2$ at the Fermi energy, $k_F$ is defined in Eq.~(\ref{k_F}), and $\mu_{s's}(q)$ is given in Eq.~(\ref{A:mu}).
After the integration over $p$ in Eq.~(\ref{A:fp3D}) (e.g., integration by parts), we substitute $f_{s's}(q)$ into Eq.~(\ref{A:chif}), which gives us Eq.~(\ref{chi03D}).

\section{Dimensional reduction of the susceptibility diagrams}
\label{app:dimred}

Here we show how to reduce Eqs.~(\ref{chiabssprime}) and (\ref{chicssprime}) to the form of Eq.~(\ref{chired}) with effective one-dimensional susceptibilities given by Eqs.~(\ref{chiab1D}) and (\ref{chic1D}).
For this, we refer to the Appendix B in Ref.~\cite{mis2023} where the leading large-distance asymptotics of the following integral is derived
\begin{eqnarray}
&& \hspace{-10pt} \int\limits_0^{2\pi} d \phi_1 \, f(\bm r_1, \bm r_2) e^{i \nu Q |\bm r_1 - \bm r_2|} \approx \sum\limits_{\sigma_1 = \pm 1} f(\sigma_1 r_1 \bm n_2, \bm r_2)  \nonumber \\
&& \hspace{20pt} \times e^{i \nu Q |\sigma_1 r_1 - r_2|} e^{i \nu \sigma_1 \frac{\pi}{4}} \sqrt{\frac{2 \pi |\sigma_1 r_1 - r_2|}{Q r_1 r_2}} , \label{B:angint}
\end{eqnarray}
where $f(\bm r_1, \bm r_2)$ is an arbitrary function that changes slowly at $Q r_1 \gg 1$, $Q r_2 \gg 1$, the integration is taken over the angle $\phi_1$ between 2D vectors $\bm r_1$ and $\bm r_2$, such that $r_1 = |\bm r_1|$ and $\bm r_2$ are fixed, $\nu = \pm 1$, and $\bm n_2 = \bm r_2/r_2$ is the unit vector along $\bm r_2$.
The asymptotic expansion given by Eq.~(\ref{B:angint}) follows from the stationary phase method. The stationary points correspond to the extrema of $|\bm r_1 - \bm r_2| = \sqrt{r_1^2 + r_2^2 - 2 r_1 r_2 \cos(\phi_1)}$, i.e. the leading large-distance contribution to the integral comes from the directions of $\bm r_1$ that are nearly collinear to $\bm r_2$.

Let us begin by integrating out the angular dimensions in Eq.~(\ref{chiabssprime}) starting from $\bm r_2$.
For brevity, we only show the part of Eq.~(\ref{chiabssprime}) that depends on $\bm r_2$, the rest is denoted by dots,
\begin{eqnarray}
&& \hspace{-15pt} \chi_{s's}^a(\bm r, \tau) \nonumber \\
&& = \int \dots \int d\bm r_2 \, V(\bm r_1 - \bm r_2) G_{s'} (\bm r_1 - \bm r_2) G_{s'}(r_2) ,
\end{eqnarray}
where we also suppressed the time arguments for brevity of expressions as the time integrations are unaffected by this procedure.
Substituting the large-distance asymptotics of the Green's function, see Eq.~(\ref{eq:green}), we find
\begin{eqnarray}
&& \hspace{-0pt} \chi_{s's}^a(\bm r, \tau) \approx \! \int \! \dots \! \int\limits_0^\infty dr_2 \, r_2 \! \int\limits_0^{2\pi} \! d\phi_2 \sum\limits_{\nu_1, \nu_2} \! \frac{e^{-i \frac{\pi}{4}(\nu_1 + \nu_2)}g_{s'}^{\nu_2}(r_2)}{\lambda_{s'} \sqrt{r_2 |\bm r_1 - \bm r_2|}} \nonumber \\
&& \hspace{-0pt} \times  V(\bm r_1 - \bm r_2) g_{s'}^{\nu_1}(|\bm r_1 - \bm r_2|)  e^{i k_{s'}(\nu_1 |\bm r_1 - \bm r_2| + \nu_2 r_2)}  , \label{B:chia1}
\end{eqnarray}
where $\phi_2$ is the angle between $\bm r_2$ and $\bm r_1$ at fixed $\bm r_1$.
As $V(\bm r_1 - \bm r_2)$, $|\bm r_1 - \bm r_2|^{-1/2}$, and $g_{s'}^{\nu_1}(|\bm r_1 - \bm r_2|)$ change slowly on the scale of Fermi wavelength at $k_F r_1 \gg 1$ and $k_F r_2 \gg 1$,
we can use Eq.~(\ref{B:angint}) with $Q = k_{s'}$ and $\nu = \nu_1$ to evaluate the leading asymptotics of the integral over $\phi_2$,
\begin{eqnarray}
&& \hspace{-6pt} \chi_{s's}^a(\bm r, \tau) \approx \int \dots \!\! \sum\limits_{\nu_1, \nu_2, \sigma_1} \!\!\! \frac{e^{-i \frac{\pi}{4}(\nu_1 (1 - \sigma_1) + \nu_2)}}{\sqrt{\lambda_{s'} r_1}} \int\limits_0^\infty dr_2 \, g_{s'}^{\nu_2}(r_2)  \nonumber \\
&& \hspace{-6pt} \times V(|r_1 - \sigma_1 r_2|) g_{s'}^{\nu_1}(|r_1 \!- \!\sigma_1 r_2|) e^{i k_{s'}(\nu_1 |r_1\! -\! \sigma_1 r_2| + \nu_2 r_2)}  . \label{B:chia2}
\end{eqnarray}
Next, we sum over $\nu_1 = \pm 1$ and only take into account the sector where the oscillatory phase in Eq.~(\ref{B:chia2}) is independent of $r_2$. This corresponds to the following choice of $\nu_1$
\begin{eqnarray}
&& \nu_1 = \sigma_1 \nu_2 \sgn(r_1 - \sigma_1 r_2) , \label{B:nu1} \\
&& g_{s'}^{\nu_1}(|r_1 - \sigma_1 r_2|) = g_{s'}^{\sigma_1 \nu_2}(r_1 - \sigma_1 r_2) , \\
&& g_{s'}^{\nu_2}(r_2) = g_{s'}^{\sigma_1 \nu_2}(\sigma_1 r_2) , \\
&& \nu_1 (1 - \sigma_1) + \nu_2 = \sigma_1 \nu_2 , \\
&& \nu_1 |r_1 - \sigma_1 r_2| + \nu_2 r_2 = \sigma_1 \nu_2 r_1 , 
\end{eqnarray}
where $\sgn(x)$ returns the sign of $x$, here we also simplified different terms in Eq.~(\ref{B:chia2}) at $\nu_1 = \sigma_1 \nu_2 \sgn(r_1 - \sigma_1 r_2)$. 
At this point we neglect the contribution coming from $\nu_1 = - \sigma_1 \nu_2 \sgn(r_1 - \sigma_1 r_2)$ as there is the oscillatory factor $e^{2 i k_{s'} \nu_2 r_2}$ that results in very fast convergence of the integral over $r_2$ on the scale of the Fermi wavelength.
Thus, the leading term in Eq.~(\ref{B:chia2}) is the following
\begin{eqnarray}
&& \hspace{-22pt} \chi_{s's}^a(\bm r, \tau) \approx \int \dots \sum\limits_{\nu_2, \sigma_1} \frac{e^{- i \frac{\pi}{4} \nu_2}}{\sqrt{\lambda_{s'} r_1}} \int\limits_0^\infty dr_2 \, g_{s'}^{\nu_2}(\sigma_1 r_2) \nonumber\\
&& \hspace{20pt} \times V(|r_1 - \sigma_1 r_2|) g_{s'}^{\nu_2}(r_1 - \sigma_1 r_2) e^{i k_{s'} \nu_2 r_1} , \label{B:chia3}
\end{eqnarray}
where we relabeled $\sigma_1 \nu_2 \to \nu_2$.
As the index $\sigma_1$ in Eq.~(\ref{B:chia3}) is only present in the combination $\sigma_1 r_2$, the summation over $\sigma_1$ just extends the integral over $r_2$ to the real line. We thus relabel this coordinate to $x_2$,
\begin{eqnarray}
&& \hspace{-20pt} \chi_{s's}^a(\bm r, \tau) \approx \int \dots \sum\limits_{\nu_2} \frac{e^{- i \frac{\pi}{4} \nu_2}}{\sqrt{\lambda_{s'} r_1}} \int\limits_{-\infty}^\infty dx_2 \, g_{s'}^{\nu_2}(x_2) \nonumber\\
&& \hspace{22pt} \times V(|r_1 - x_2|) g_{s'}^{\nu_2}(r_1 - x_2) e^{i k_{s'} \nu_2 r_1} . \label{B:chia4}
\end{eqnarray}
Now we also restore the terms in Eq.~(\ref{B:chia4}) that depend on $\bm r_1$,
\begin{eqnarray}
&& \hspace{0pt} \chi_{s's}^a(\bm r, \tau) \approx \int \dots \int d \bm r_1 \, G_{s'}(\bm r - \bm r_1) \sum\limits_{\nu_2} \frac{e^{- i \frac{\pi}{4} \nu_2}}{\sqrt{\lambda_{s'} r_1}} e^{i k_{s'} \nu_2 r_1}  \nonumber\\
&& \hspace{30pt} \times \int\limits_{-\infty}^\infty dx_2 \, V(|r_1 - x_2|) g_{s'}^{\nu_2}(r_1 - x_2) g_{s'}^{\nu_2}(x_2) . \label{B:chia5}
\end{eqnarray}
Substituting the asymptotics of $G_{s'}(\bm r - \bm r_1)$ [see Eq.~(\ref{eq:green})] into Eq.~(\ref{B:chia5}) and integrating over $\phi_1$, the angle between $\bm r_1$ and $\bm r$, using the asymptotic relation Eq.~(\ref{B:angint}), we find
\begin{eqnarray}
&& \hspace{-5pt} \chi_{s's}^a(\bm r, \tau) \approx \int \dots \int\limits_0^\infty dr_1 \, \sum\limits_{\nu_1, \nu_2, \sigma_1} \frac{e^{- i \frac{\pi}{4}(\nu_1 (1 - \sigma_1) + \nu_2)}}{\sqrt{\lambda_{s'} r}}  \nonumber \\
&& \hspace{-5pt} \times \int\limits_{-\infty}^\infty dx_2 \,  V(|r_1 - x_2|) g_{s'}^{\nu_1}(|r - \sigma_1 r_1|) g_{s'}^{\nu_2}(r_1 - x_2) g_{s'}^{\nu_2}(x_2) \nonumber \\
&& \hspace{35pt} \times e^{i k_{s'} (\nu_1 |r - \sigma_1 r_1| + \nu_2 r_1)} . \label{B:chia6}
\end{eqnarray}
The non-oscillatory sector in Eq.~(\ref{B:chia6}) corresponds to $\nu_1 = \sigma_1 \nu_2 \sgn(r - \sigma_1 r_1)$, only this sector is important for the leading large-distance asymptotics of $\chi_{s's}(\bm r, \tau)$.
Relabeling $\sigma_1 \nu_2 \to \nu_2$ and $x_2 \to \sigma_1 x_2$ and then summing over $\sigma_1$ gives us the following result
\begin{eqnarray}
&& \hspace{-3pt} \chi_{s's}^a (\bm r, \tau) \approx \mathfrak{g} \, G_{s}(-\bm r) \sum\limits_{\nu_2} \frac{e^{i \nu_2 \left(k_{s'} r - \frac{\pi}{4} \right)}}{\sqrt{\lambda_{s'} r}} \int d\tau_1 d\tau_2 \int\limits_{-\infty}^\infty dx_1 \nonumber \\
&& \hspace{-3pt} \int\limits_{-\infty}^\infty dx_2 \, V(x_1 - x_2) g_{s'}^{\nu_2}(r - x_1) g_{s'}^{\nu_2}(x_1 - x_2) g_{s'}^{\nu_2}(x_2) . \label{B:chiafin}
\end{eqnarray}
Finally, substituting the asymptotics of $G_s(-\bm r)$ and restoring the time arguments,
we find that $\chi_{s's}^a(\bm r, \tau)$ can be represented in the form of Eq.~(\ref{chired}) with the one-dimensional susceptibility $\tilde \chi^a_{\gamma' \gamma}(r, \tau)$ given by Eq.~(\ref{chiab1D}).

The dimensional reduction procedure can be applied exactly the same way to Eq.~(\ref{chicssprime}).

\section{One-dimensional static susceptibilities}
\label{app:1D}

First, we derive Eqs.~(\ref{chiabstatic1D}), (\ref{chicstatic1D}) for the one-dimensional static susceptibilities.
For this, we only have to evaluate integrals over $\tau$ and $\tau_1$, see Eqs.~(\ref{chiab1D}), (\ref{chic1D}), (\ref{static1D}), the integral over $\tau_2$ is trivial for instantaneous interactions, see Eq.~(\ref{Vdelta}).
Integrals over imaginary times at zero temperature are simple-pole integrals, for example,
\begin{eqnarray}
&& \hspace{-20pt} \int\limits_{-\infty}^\infty d\tau \, g_\gamma (-x, - \tau) g_{\gamma'}(x - x_1, \tau - \tau_1)  \nonumber \\
&& = - \frac{1}{2 \pi} \frac{\vartheta\left(-\nu \nu' x (x - x_1)\right)}{v_{s'} |x| + v_s |x - x_1| + i \nu v_{s} v_{s'} \tau_1 \sgn(x)} , \label{C:t1} 
\end{eqnarray}
where $g_\gamma(x, \tau)$ is defined by Eqs.~(\ref{gs}), (\ref{ggamma}), $\gamma = \{\nu, s\}$, $\gamma' = \{\nu, s'\}$, and $\sgn(x)$ returns the sign of $x$.
In fact, all other time integrals can be expressed through the integral in Eq.~(\ref{C:t1}) by relabeling the arguments.
Performing the time integrals using Eq.~(\ref{C:t1}), we get Eqs.~(\ref{chiabstatic1D}), (\ref{chicstatic1D}).

Next, we derive Eq.~(\ref{chisumr}) starting from Eqs.~(\ref{chiabstatic1D}), (\ref{chicstatic1D}) and (\ref{chisum}).
Relabeling $|x| \to r$ in Eq.~(\ref{chiabstatic1D}) and then making change of variables $x_2 - x_1 = z$, $x_2 = y$, we find
\begin{eqnarray}
&& \hspace{-25pt} \tilde \chi^a \left(Q_{\gamma' \gamma}, r\right) \nonumber \\
&& = - \frac{1}{4 \pi^2 v_s v_{s'}} \int\limits_{-\sigma r}^{+\infty} dz\, \frac{V(z)}{z} \frac{z + \sigma r}{z + \left(\frac{v_{s'}}{v_s} + \sigma\right) r} , \label{C:ab}
\end{eqnarray}
where $\sigma = - \nu \nu'$. Here we already accounted for the fact that the integration is taken over the region $\{(x_1, x_2) \! : \, x_1 < \sigma r, \, x_2 > 0 \} = \{(y, z) \! : z > -\sigma r, \, z + \sigma r > y > 0 \}$.
As the function under the integral depends only on $z$, the integration over $y$ just yields the factor $z + \sigma r$ in Eq.~(\ref{C:ab}).
Notice that here $z$ is just a real variable.

In Eq.~(\ref{chicstatic1D}) we relabel $|x| \to r$ and make the change of variables $z = x_2 - x_1$, $y = x_1 + \sigma x_2 v_{s'}/v_s$, $\sigma = - \nu_1 \nu_2$.
In the case $\sigma = +1$, the integration is taken over the region $\mathcal{R}^+ = \mathcal{R}_1^+ \cup \mathcal{R}_2^+ \cup \mathcal{R}_3^+$ consisting of three square regions: $\mathcal{R}_1^+ = \{(x_1, x_2) \!: \, x_1 > r, \, x_2 > r \}$, $\mathcal{R}_2^+ = \{(x_1, x_2) \!: \, r > x_1 > 0, \, r > x_2 > 0 \}$, and $\mathcal{R}_3^+ = \{(x_1, x_2) \!: \, x_1 < 0, \, x_2 < 0 \}$.
In case $\sigma = -1$, the integration region $\mathcal{R}^- = \mathcal{R}^-_{1} \cup \mathcal{R}^-_{2}$ consists of two square regions: $\mathcal{R}_1^- = \{(x_1, x_2) \!: \, x_1 < 0, \, x_2 > r \}$, and $\mathcal{R}_2^- = \{(x_1, x_2) \!: \, x_1 > r, \, x_2 < 0 \}$.
The integration over $y$ can be performed in all five different cases,
\begin{eqnarray}
&& \hspace{-10pt} \tilde \chi^c\left(Q_{\gamma' \gamma}, r| \mathcal{R}_1^+\right) = \tilde \chi^c\left(Q_{\gamma' \gamma}, r| \mathcal{R}_3^+\right) \nonumber \\
&& \hspace{10pt} = \frac{1}{v_s v_{s'} r a_+} \int\limits_0^{+\infty} \ln\left|1 + a_+ \frac{r (r + z)}{z^2}\right| \frac{V(z) \, dz}{4 \pi^2} , \label{C:R1plus} \\
&& \hspace{-10pt} \tilde \chi^c \left(Q_{\gamma' \gamma}, r| \mathcal{R}_2^+\right) \nonumber \\
&& \hspace{10pt} = \frac{2}{v_s v_{s'} r a_+} \int\limits_0^{r} \ln\left|1 + a_+ \frac{r (r - z)}{z^2}\right| \frac{V(z) \, dz}{4 \pi^2} , \label{C:R2plus}  \\
&& \hspace{-10pt} \tilde \chi^c \left(Q_{\gamma' \gamma}, r| \mathcal{R}_1^-\right) = \tilde \chi^c \left(Q_{\gamma' \gamma}, r| \mathcal{R}_2^-\right) \nonumber \\
&& \hspace{10pt} = \frac{1}{v_s v_{s'} r a_-} \int\limits_r^{+\infty} \ln\left|1 + a_- \frac{r (z - r)}{z^2}\right| \frac{V(z) \, dz}{4 \pi^2} , \label{C:R1minus} 
\end{eqnarray}
where $a_\pm$ is given by Eq.~(\ref{a}).
Combining all $\mathcal{R}_i^\sigma$ contributions for a given $\sigma$, we can represent $\tilde \chi^c \left(Q_{\gamma' \gamma}, r \right)$ in the following form
\begin{eqnarray}
&& \hspace{-25pt}  \tilde \chi^c \left(Q_{\gamma' \gamma}, r\right) \nonumber \\
&& \hspace{10pt} = \frac{1}{r a_\sigma} \int\limits_{-\sigma r}^{+\infty} \ln\left|1 + a_\sigma \frac{r (z + \sigma r)}{z^2}\right| \frac{V(z) \, dz}{2 \pi^2 v_s v_{s'}} , \label{C:c}
\end{eqnarray}
where $\sigma = -\nu \nu'$ and $a_\sigma$ is given by Eq.~(\ref{a}).
Combining Eqs.~(\ref{C:ab}), (\ref{C:c}), we get Eq.~(\ref{chisumr}).

\section{Useful Fourier transforms}
\label{app:Fourier}

First, we use the asymptotics of the following $D$-dimensional angular integral at $q r \gg 1$
\begin{eqnarray}
&& \hspace{-20pt} \int d \bm n_{\bm r} \, e^{- i \bm q \cdot \bm r} \approx \left(\frac{2 \pi}{q r}\right)^{\frac{D - 1}{2}} 2 \cos \left(q r - \frac{\pi}{4}(D - 1)\right) , \label{D:ang}
\end{eqnarray}
where $\bm n_{\bm r} = \bm r/r$, and $d \bm n_{\bm r}$ is a surface element of the unit $(D - 1)$-dimensional sphere.
This allows us to find the asymptotic behavior of the following $D$-dimensional Fourier transform at $q - Q \ll Q$, 
\begin{eqnarray}
&& \int \! d^D \bm r \,  \frac{e^{i \nu Q r}}{r^{\frac{D - 1}{2}}} f(r) e^{- i \bm q \cdot \bm r} \! = \! \int\limits_0^\infty \! dr \, f(r) r^{\frac{D - 1}{2}}  e^{i \nu Q r} \!\!\! \int \!\! d \bm n_{\bm r} \, e^{- i \bm q \cdot \bm r} \nonumber \\
&& \approx \left|\frac{2 \pi}{Q}\right|^{\frac{D - 1}{2}} e^{i \nu\frac{\pi}{4}(D - 1)} \int\limits_0^{+\infty} dr \, f(r) e^{- i \nu (q - Q) r} , \label{D:Q}
\end{eqnarray}
where $f(r)$ is an arbitrary function that changes slowly on the scale of $1/ Q$ at $Q r \gg 1$, $\nu = \pm 1$.
We use Eq.~(\ref{D:Q}) to calculate the non-analytic Kohn anomalies in momentum space.

In this paper we also make use of the following integrals (and their analytic continuations)
\begin{eqnarray}
&& \hspace{-20pt} \mathcal{L}_\nu (\alpha, q) \equiv \int\limits_0^\infty \frac{dr}{r^{\alpha}} \, e^{- i \nu q r} = \frac{\pi |q|^{\alpha - 1} e^{i \nu \sgn(q) \frac{\pi}{2}(\alpha - 1)}}{\sin(\pi \alpha) \Gamma(\alpha)}  ,  \label{D:L} \\
&& \hspace{-20pt} \int\limits_0^\infty \frac{dr}{r^{\alpha}} \, e^{- i \nu q r} \left[\ln(r)\right]^{n} = \left(-\frac{\partial}{\partial \alpha}\right)^n \mathcal{L}_\nu(\alpha, q) , \label{D:log}
\end{eqnarray}
where $\nu = \pm 1$, $q$ is a real variable, $\Gamma(x)$ is the Euler gamma function, $n$ is a positive integer, and $\alpha$ is an arbitrary real number.
We often use Eq.~(\ref{D:L}) for $\alpha \ge 1$ where the integral $\mathcal{L}(\alpha, q)$ is formally divergent.
In our case this means that a short-distance cut-off must be introduced.
However, the non-analyticities/discontinuities of susceptibilities can be found exactly via the analytic continuation of this divergent integral which is given by the right-hand side of Eq.~(\ref{D:L}).

\section{$V_1$ and $V_2$ for the Thomas-Fermi interaction}
\label{app:TF}

First, we prove the representations of $V_1$ and $V_2$ given in Eqs.~(\ref{V1}), (\ref{V2}).
For this, we use the 2D Fourier transform of $V_{R_0}(x)$
\begin{eqnarray}
&& V_{R_0}(x) = \int\limits_0^\infty \frac{dq}{2\pi} \, q J_0(q x) V_{R_0}(q) , \label{E:Fourier}
\end{eqnarray}
where $J_0(x)$ is the Bessel function, and $V_{R_0}(q)$ is the 2D Fourier transform of $V_{R_0}(x)$.
Substituting Eq.~(\ref{E:Fourier}) into Eqs.~(\ref{V1}), (\ref{V2}), we find
\begin{eqnarray}
&& V_1 = - \int\limits_0^\infty \frac{dq}{2 \pi} \, V_{R_0} (q) \int\limits_0^\infty dz \, J_0(z) \ln\left|\frac{z}{q R_0}\right| , \label{E:V1}\\
&& V_2 = \int\limits_0^\infty \frac{dq}{2 \pi} \, V_{R_0} (q) \int\limits_0^\infty dz \, J_0(z) , \label{E:V2}
\end{eqnarray}
where we introduced the new integration variable $z = q x$.
$V_1$ and $V_2$ then follow from the following identities
\begin{eqnarray}
&& \int\limits_0^\infty dz \, J_0(z) \ln(z) = - \gamma - \ln 2 , \\
&& \int\limits_0^\infty dz \, J_0(z) = 1 ,
\end{eqnarray}
where, again, $\gamma$ is the Euler-Mascheroni constant.

Substituting $V_{TF}(q)$ [see Eq.~(\ref{VTFq})] into Eqs.~(\ref{V1}), (\ref{V2}), we find
\begin{eqnarray}
&& V_1^{TF} = \int\limits_0^{k_\Lambda} \frac{dq}{m a_B} \, \frac{\ln(2 q R_0) + \gamma}{q + \kappa} , \label{E:V1TFq} \\
&& V_2^{TF} = \int\limits_0^{k_\Lambda} \frac{dq}{m a_B} \, \frac{1}{q + \kappa} \approx \frac{1}{m a_B} \ln\left(\frac{k_\Lambda}{\kappa}\right) , \label{E:V2TF}
\end{eqnarray}
where we introduced the ultraviolet cut-off $k_\Lambda \sim k_F \gg \kappa$.
Here, Eq.~(\ref{E:V2TF}) already coincides with Eq.~(\ref{V2TF}).
In order to get Eq.~(\ref{V1TF}), we use the following identity
\begin{eqnarray}
&& \int\limits_0^1 \frac{\ln u \, du}{u + z} = \frac{\left[\ln (1 + z)\right]^2 -\left[ \ln z\right]^2}{2} - \frac{\pi^2}{6} \nonumber \\
&& \hspace{46pt} + {\rm Li}_2\left(\frac{z}{1 + z}\right) , \label{E:polylog}
\end{eqnarray}
where $z > 0$, and ${\rm Li}_2(z) = \sum_{n = 1}^\infty z^n/n^2$ is the polylogarithm.
Rescaling the integration variable $q = k_\Lambda u$ in Eq.~(\ref{E:V1TFq}) results in an integral like Eq.~(\ref{E:polylog}) with $z = \kappa/k_\Lambda$, plus a simple logarithmic integral like Eq.~(\ref{E:V2TF}).
As $\kappa \ll k_\Lambda$, we then expand Eq.~(\ref{E:polylog}) keeping only the leading logarithmic order, which gives us Eq.~(\ref{V1TF}).

\bibliographystyle{apsrev4-2}
\bibliography{bibTMD}
\end{document}